\def\mearth{{\rm\,M_\oplus}}
\def\mjup{{\rm\,M_{Jup}}}
\def\vinf{{v_{\infty}}}

\def\msun{{\rm\,M_\odot}}
\def\deg{^\circ}

\def\Chi{\raisebox{2pt}{$\chi$}}

\documentclass[preprint,authoryear,12pt]{elsarticle}
\usepackage{amssymb}
\usepackage{aas_macros}

\begin{document}
 
\begin{frontmatter}

\title{Was the Solar System's dynamical instability triggered by a (sub)stellar flyby?} 

\author{Sean N. Raymond$^{a}$ \& Nathan A. Kaib$^{b}$\\
\address[1]{Laboratoire d'Astrophysique de Bordeaux, Univ. Bordeaux, CNRS, B18N, All{\'e}e Geoffroy Saint-Hilaire, 33615 Pessac, France; rayray.sean@gmail.com}
\address[2]{Planetary Science Institute, 1700 E. Fort Lowell, Suite 106, Tucson, AZ 85719, USA}
}


\begin{abstract}

An instability among the giant planets' orbits can match many aspects of the Solar System's current orbital architecture. We explore the possibility that this dynamical instability was triggered by the close passage of a star or substellar object during the Sun's embedded cluster phase. We run N-body simulations starting with the giant planets in a resonant chain and an outer planetesimal disk, with a wide-enough planet-disk separation to preserve the planets' orbital stability for $>$100 Myr. We subject the system to a single flyby, testing a wide range in flyby mass, velocity and closest approach distance. We find a variety of outcomes, from flybys that over-excite the system (or strip the planets entirely) to flybys too weak to perturb the planets at all.  An intermediate range of flybys triggers a dynamical instability that matches the present-day Solar System. Successful simulations -- that match the giant planets' orbits without over-exciting the cold classical Kuiper belt -- are characterized by the flyby of a substellar object ($3-30 \mjup$) passing within 20 au of the Sun. We performed Monte Carlo simulations of the Sun's birth cluster phase, parameterized by the product of the stellar density $\eta$ and the cluster lifetime $T$. The balance between under- and over-excitation of the young Solar System is at $\eta T \approx 5 \times 10^4$~Myr pc$^{-3}$, in a range consistent with previous work. We find a probability of $\sim$1\% that the Solar System's dynamical instability was triggered by a substellar flyby.  The probability increases to $\sim$5\% if the occurrence rate of free-floating planets and low-mass brown dwarfs is modestly higher than predicted by standard stellar initial mass functions. 
\end{abstract}


\end{frontmatter}


\section{Introduction}
The giant planet dynamical instability model for the Solar System was developed as a confluence of several different fields of study.  \cite{fernandez84} showed that interactions between the planets and a planetesimal disk would invariably lead to a spreading out of the giant planets' orbits.  The ice giants' and Saturn's orbits expand due to preferentially scattering planetesimals inward, whereas Jupiter's orbit contracts due to preferentially ejecting them~\citep[see also][]{hahn99,gomes04}.  \cite{malhotra93,malhotra95} showed that Neptune's outward migration could explain the 2:3 resonant orbit of Pluto. Meanwhile, studies showed that the eccentric orbits of giant exoplanets could be explained by dynamical instabilities in which strong gravitational kicks typically lead to the ejection of one or more planets~\citep{weidenschilling96,rasio96,lin97}. \cite{thommes99,thommes02} proposed that a drastic re-arrangement of the Solar System's giant planets took place after their formation, by invoking that the ice giants accreted in the Jupiter-Saturn region and then were scattered outward and captured by interactions with a massive outer planetesimal disk. Part of the motivation of this scenario came from the fact that planetesimal accretion could not plausibly form the ice giants within a reasonable timeframe, or really at all at their current locations, as the growing ice giants will instead scatter planetesimals inward toward the gas giants once their escape speeds exceeded their orbital velocities~\citep[e.g.][]{thommes03,levison10}.  This growth constraint was lifted with the advent of pebble accretion~\citep{ormel10,lambrechts12}.  The so-called `Nice model' proposed that the giant planets formed in a more compact orbital configuration that became unstable through interactions with an outer planetesimal disk~\citep{tsiganis05,morby05,gomes05}. The initial conditions of the early variant were ad hoc, but later incarnations invoked orbital configurations emerging from simulations of gas-driven orbital migration~\citep{morby07}.  While some details have changed, the Nice model forms of the basis of the current paradigm.

The instability model can reproduce a host of characteristics of the present-day Solar System~\citep[for a review, see][]{nesvorny18b}. It can match the broad orbital inclination distribution of Jupiter's co-orbital asteroids~\citep[which were captured during the instability;][]{morby05}, the existence and orbital distribution of the giant planets' irregular satellites~\citep[also captured during the instability; ][]{nesvorny07}, and the photometric match between these populations and Kuiper belt objects~\citep{morby09c}. The orbital structure of the Kuiper belt can be explained by Neptune's outward migration in the aftermath of the instability phase~\citep[e.g.][]{wolff12,dawson12,nesvorny15,nesvorny18b}, as can certain aspects of the orbital structure of the asteroid belt~\citep[e.g.][]{minton09,clement20}. Simulations of the instability also match the orbits of the giant planets~\citep{tsiganis05,nesvorny12,batygin12b}, albeit with two important caveats. First, the young Solar System likely included a third ice giant~\citep{nesvorny11}. The instability almost always leads to the ejection of an ice giant, such that the odds of matching the giant planets' orbits increase drastically if this additional planet's existence is invoked.  Second, while the instability adequately matches the giant planets' semimajor axes and inclinations, it struggles to excite Jupiter's eccentricity sufficiently.  To date, the only plausible solution to this issue is that Jupiter's eccentricity was already non-zero before the instability (Pierens et al 2014; Clement et al 2021a, 2021b). 

The Nice model instability was conceived as a delayed event~\citep{gomes05}, to match the `terminal lunar cataclysm', a perceived spike in lunar bombardment flux $\sim 600$~Myr after the start of planet formation~\citep{tera74}.  However, recent re-analysis of the lunar cratering record -- as well as the demonstration that a late instability would almost certainly have destroyed our system of terrestrial planets~\citep{kaib16} -- has upended the idea of a terminal lunar cataclysm~\citep{boehnke16,zellner17,hartmann19}. Rather, it appears likely that the Moon's bombardment rate has declined steadily, and that the spike in ages found by Apollo samples were caused by selection bias. Given the number of features of the Solar System that can be explained by the instability, it is still likely to have taken place, but its timing is not anchored to the lunar cratering record.

The dynamical instability must have taken place early, but the exact timing is hard to constrain precisely. \cite{morby18} used the concentrations of highly-siderophile elements in Earth and the Moon, as well as the lunar cratering record, to argue that the instability took place no later than 100 Myr after the start of planet formation~\citep[as marked by the ages of CAIs; ][]{bouvier10,connelly12}.  \cite{nesvorny18} showed that an instability within $\sim 20-100$~Myr is required to preserve the known binary Jupiter Trojan. \cite{mojzsis19} used reset ages of meteorites to argue that the instability could not have happened later than $\sim 50$ Myr after CAIs. \cite{harper_edwards24} used the K-Ar age distribution of asteroidal meteorites to argue the instability must have happened earlier still, likely between 5 and 20~Myr after CAIs. A similar, very early timeframe was also inferred by \cite{hunt22} using the age distribution of iron meteorites.  A very early instability (no later than 10-20 Myr after CAIs) thus appears likely based on current empirical constraints.\footnote{\cite{avdellidou24} used a chain of reasoning related to the collisional and orbital history of the asteroid Athor's parent body to argue that the instability must have instead taken place significantly later, between 60 and 100 Myr after CAIs. However, \cite{izidoro25b} showed that a key assumption made by \cite{avdellidou24} related to the dynamical implantation of asteroids relative to the timing of instability is incorrect, invalidating their timeline.}

The instability timing is inextricably linked with its trigger.  In systems without outer planetesimal disks, there is a broad range in the timescale for the onset of instability, with a characteristic range of $\sim 10^{4-6}$ years, depending on the planets' initial spacing and masses~\citep{marzari02,chatterjee08,raymond10}. Yet the early Solar System harbored a massive outer planetesimal disk, which may have played a role.  To date, three distinct mechanisms have been proposed as instability triggers in the Solar System (ordered roughly by time to onset of instability): 
\begin{enumerate}
\item Gas disk-driven instability. If the late evolution of the Sun's protoplanetary disk was dominated by a photo-evaporative wind driven by irradiation from the central star, the disk would have dispersed from the inside-out~\citep[see review in ][]{alexander14}. The inner edge of the outer gas disk would have swept outward. \cite{liu22} showed that this outward-sweeping edge causes a compression of the giant planet system and invariably triggers dynamical instability.  In this model, the instability would have been causally linked with the disk's dispersal, which likely took place $\sim 5$~Myr after CAIs~\citep{wang17b,hunt22}.
\item Self-driven instability.  The giant planets may have emerged from the protoplanetary disk in a configuration that was simply not long-term stable (without the disk's dispersal being responsible for triggering the instability). Using simulations of the giant planets interacting with a planetesimal disk that was sculpted by the ice giants' growth, \cite{ribeiro20} showed that about half of dynamical instabilities were triggered solely by planet-planet interactions, on a characteristic timescale of a few Myr after disk dispersal.  Hydrodynamic simulations of giant planet growth and migration within the gaseous disk, followed by its dispersal, produce similar outcomes~\citep[e.g.][]{griveaud24}.
\item Planetesimal disk-driven instability. The original Nice model instability trigger involved slow spreading of the giant planets' orbits until Jupiter and Saturn crossed the 2:1 resonance, which caused a sudden increase in their eccentricities and triggered a global instability~\citep{tsiganis05,morby07}. In that setup, the instability timescale scaled with the inner edge of the planetesimal disk~\citep{gomes05}. Using self-consistently sculpted planetesimal disks, \cite{ribeiro20} found that instabilities driven by planet-planetesimal disk interactions tend to have timescales of 30-60 Myr.  Simulations that include the self-gravity of the planetesimal disk, but not its primordial sculpting during the ice giants' growth, find a similar instability timescale~\citep{levison10,quarles19,kaib24b}. 
\end{enumerate}

In this paper, we explore the possibility that a stellar flyby during the Sun's birth cluster phase could have triggered the giant planet dynamical instability.  The Sun formed in a cluster with between a few hundred and roughly ten thousand members~\citep[e.g.,][]{adams01,portegieszwart09,adams10,pfalzner13,portegieszwart19}.  At least a few hundred stars are needed in order for the star-forming region to contain a single star massive enough to have produced and injected the Aluminum-26 thought to have been responsible for widespread melting of planetesimals in early Solar System history~\citep[e.g.,][]{hester04,gaidos09,gounelle12,monteux18,arakawa23}. With more than about 10,000 stars, the odds of a close encounter that would disrupt the observed dynamically cold orbits of the cold classical Kuiper belt~\citep{brown01,petit11,gladman21} increase dramatically~\citep[e.g.][]{kobayashi01,adams10}.  More precisely, the odds of disrupting the cold classicals scales with the product of the stellar density $\eta$ and the cluster lifetime $T$~\citep{batygin20,nesvorny23}.  

Based on the dynamics of clusters in the relevant mass range, the closest stellar flyby of the Sun during this phase is thought to have been $\sim$100-200~au~\citep[e.g.][]{kobayashi01,malmberg11,arakawa23}.  Such a flyby has been invoked in previous studies to explain certain orbital characteristics of the trans-Neptunian objects, such as the existence of dynamically-detached objects and  the radial distribution of TNOs~\citep{kenyon04b,morby04,brasser06,brasser15,nesvorny23,pfalzner24}. Yet it is possible that the Sun suffered a much closer flyby. Star-forming regions contain rich sub-stellar populations that extend into the planetary mass regime~\citep[e.g.][]{scholz12,penaramirez12,miretroig22}. The impulse imparted by a flyby scales with the object's mass, such that a brown dwarf or free-floating planet can pass much closer to the planets' (or small bodies') orbits without causing significant disruption.  \cite{brown25} showed that the flyby of a $\sim 8 \mjup$ object within less than 2 au of the Sun could have excited the giant planets' orbits to their present-day secular state.  However, \cite{brown25} assumed that the giant planets started on perfectly coplanar and circular orbits at their current semimajor axes, which is at odds with models of their formation and migration within the Sun's gaseous protoplanetary disk~\citep[e.g.][]{morby07,pierens14,griveaud24}.  

Here, we present simulations of the giant planets' early dynamical evolution under the influence of a flyby. We start from a system containing the four giant planets (plus an extra ice giant) in a multi-resonant configuration and an outer planetesimal disk, with a setup carefully chosen to remain stable for at least 100 Myr if left in isolation.  We then subject this system to flybys of objects with masses as low as $1 \mjup$, while covering the relevant range in flyby velocities and close approach distances.  We apply constraints to our simulation outputs, requiring successful systems to match the giant planets' orbits and to preserve the cold classical Kuiper belt. We then put our results in the context of simple cluster models and use Monte Carlo modeling to determine the cluster parameters that are consistent with a flyby-driven dynamical instability.  We show that, for plausible clusters, there is a $\sim 10\%$ chance that the dynamical instability could have been triggered by the flyby of a brown dwarf or free-floating planet.

\section{Simulation methods}

Our simulations started with the giant planets and planetesimals orbiting the Sun, plus a flyby star or substellar object.  The giant planets started in a multi-resonant configuration taken from \cite{kaib16}, with Jupiter at 5.97 au, Saturn in 3:2 resonance at 8.01 au, the first ice giant in 2:3 resonance at 10.5 au, then the outer two pairs of ice giants in 2:1 resonance, at 15.6 au and 20.4 au.  This configuration was generated by convergent migration of the planets using synthetic forces and eccentricity damping~\citep[for details, see][]{kaib16}. The planets' orbits were perfectly coplanar. We integrated this configuration to verify that it was stable for 1 Gyr.  

Our planetary system's resilience to outside perturbation is, in part, dependent on our particular choice of resonant chain and the numerical prescription used to assemble the chain. Nonetheless, there is no reason to believe that our choice of planetary configuration is particularly fragile when compared with others in the recent literature. In about half of the simulations from \cite{ribeiro20}, the giant planets simply went unstable on their own after the gas disk dissipated.  That paper, like almost all done to date (including ours), generated chains of orbital resonances among the giant planets using N-body simulations with migration and eccentricity and inclination damping included as synthetic forces.  Yet signs point to more realistic (hydrodynamical) simulations producing chains of resonances among the giant planets that are significantly {\em less} likely to remain stable than those produced via the N-body route~\citep[e.g.][]{griveaud24}. Together, this means that giant planet systems in general likely form in configurations that make them susceptible to going unstable with only a small perturbation.  This actually fits nicely with the broad eccentricity distribution of giant exoplanets~\citep{butler06,udry07}, which indicates that the vast majority of these systems are the survivors of dynamical instability~\citep[e.g.][]{chatterjee08,juric08,raymond10}.

In addition to planets, our simulations included an outer planetesimal disk that was evenly distributed between 24 and 30 au (following an $r^{-1}$ surface density profile) and contained $20 \mearth$ evenly distributed among 1000 particles. We chose 24 au as the inner edge of the planetesimal disk to ensure that the giant planet system remained stable for at least 100 Myr in the presence of the planetesimal disk. The planetesimals' initial eccentricities were zero and initial inclinations were randomly chosen between zero and $1^\circ$.  We included two additional populations of planetesimals, which were treated as massless particles. First, the cold classical Kuiper belt was laid down with semimajor axes evenly spread between 40 and 50 au, with zero initial eccentricities and inclinations. The cold classical Kuiper belt included 200 massless particles. Second, a primordial scattered disk was laid down; such a disk is seen in the simulations of \cite{ribeiro20} of ice giant formation that also include outer planetesimals. The orbits of primordial scattered disk particles were chosen to match these simulations, with perihelion distances randomly chosen between 15 and 20 au (thus crossing the orbits of the ice giants) and semimajor axes randomly chosen up to 50 au.  The primordial scattered disk also included 200 massless particles.

Each simulation underwent a single flyby.  The object's mass was chosen using the stellar initial mass function of \cite{maschberger13}, with a mass between $1 \mjup$ and $10 \msun$. While it does not perfectly describe the mass function of the lowest-mass objects~\citep[but rather underestimates their abundance; see][]{miretroig22}, the \cite{maschberger13} IMF was chosen for its broad applicability and numerical convenience. The other flyby parameters were sampled to explore the variety of outcomes. The impact parameter $b$ was chosen randomly between 1 and 1000 au -- this serves as a test of parameter space but is not an accurate representation of real flybys, for which the probability of a given flyby scales with $b^2$. The velocity at infinity $\vinf$ was randomly chosen between 0.1 and 5 km/s, which roughly captures the velocities in embedded clusters, which are expected to have velocity dispersions $\langle v \rangle \sim 1$~km/s~\citep[e.g.][]{lada03}.  The flyby star was initialized just interior to $10^4$ au on its way toward the Sun, accounting for the fact that the star is already within the Sun's potential. The flyby typically happened within the first few tens of thousands of years of the simulation.  

We used the Bulirsch-Stoer integrator within the {\tt Mercury} integration package~\citep{chambers99}, with an accuracy parameter of $10^{-12}$ and a default (starting) timestep of 150 days.  Each simulation was first run for 10 Myr, and then continued out to 100 Myr if it had 4-5 surviving giant planets interior to 40 au and, therefore, the potential to match the Solar System (a little less than half of simulations were run out to 100 Myr).  Any object passing outside of $10^4$ au was considered to be ejected and removed from the simulation; this radius is appropriate for a relatively dense cluster~\citep{tremaine93}.  Collisions were treated as inelastic mergers conserving linear momentum.  

We ran 3000 simulations of the flyby phase and its aftermath. The first batch of 2000 simulations sampled impact parameters between 1 and 150 au.  Then, to broaden the scope of the paper, we ran a second batch of 1000 simulations with impact parameters between 150 and 1000 au.  Even though it represents a much coarser sampling of parameter space, the second batch of simulations was vital in applying our results to the broader context of the expected encounters during the stellar birth cluster phase (see Section 4).

\section{Results}

In this section, we first discuss the outcomes of the population of simulations as a whole. We then filter the results using  various success criteria related to the orbital distributions of the planets and small bodies. We also explore the dynamical evolution of successful outcomes.

One important criterion is the level of orbital excitation of the giant planet system.  This can be quantified using the normalized angular momentum deficit, a measure of the non-circularity and non-coplanarity of the planets' orbits~\citep{laskar97}.  It is defined as:
\begin{equation}
 AMD = \frac{\sum_{j} m_j \sqrt{a_j} \left(1 - cos
(i_j) \sqrt{1-e_j^2}\right)} {\sum_j m_j \sqrt{a_j}}, 
\end{equation}
\noindent where $a_j$, $e_j$, $i_j$, and $m_j$ refer to planet $j$'s semimajor axis, eccentricity, inclination with respect to a fiducial plane, and mass. The $AMD$ of the Solar System's giant planets is 0.0011, calculated with the time-averaged orbits of the giant planets from Table~1 of \cite{nesvorny12}.  The AMD of the pre-instability giant planet system in our simulations was $8 \times 10^{-5}$.

Below, we evaluate the viability of our simulated planetary systems using the $AMD$ as well as other features of the systems, such as the number of planets, their actual orbits, and the properties of small body populations (e.g., the cold classical Kuiper belt). 

\subsection{Distribution of outcomes}

\begin{figure*}

\begin{centering}
	\includegraphics[height=0.3\columnwidth]{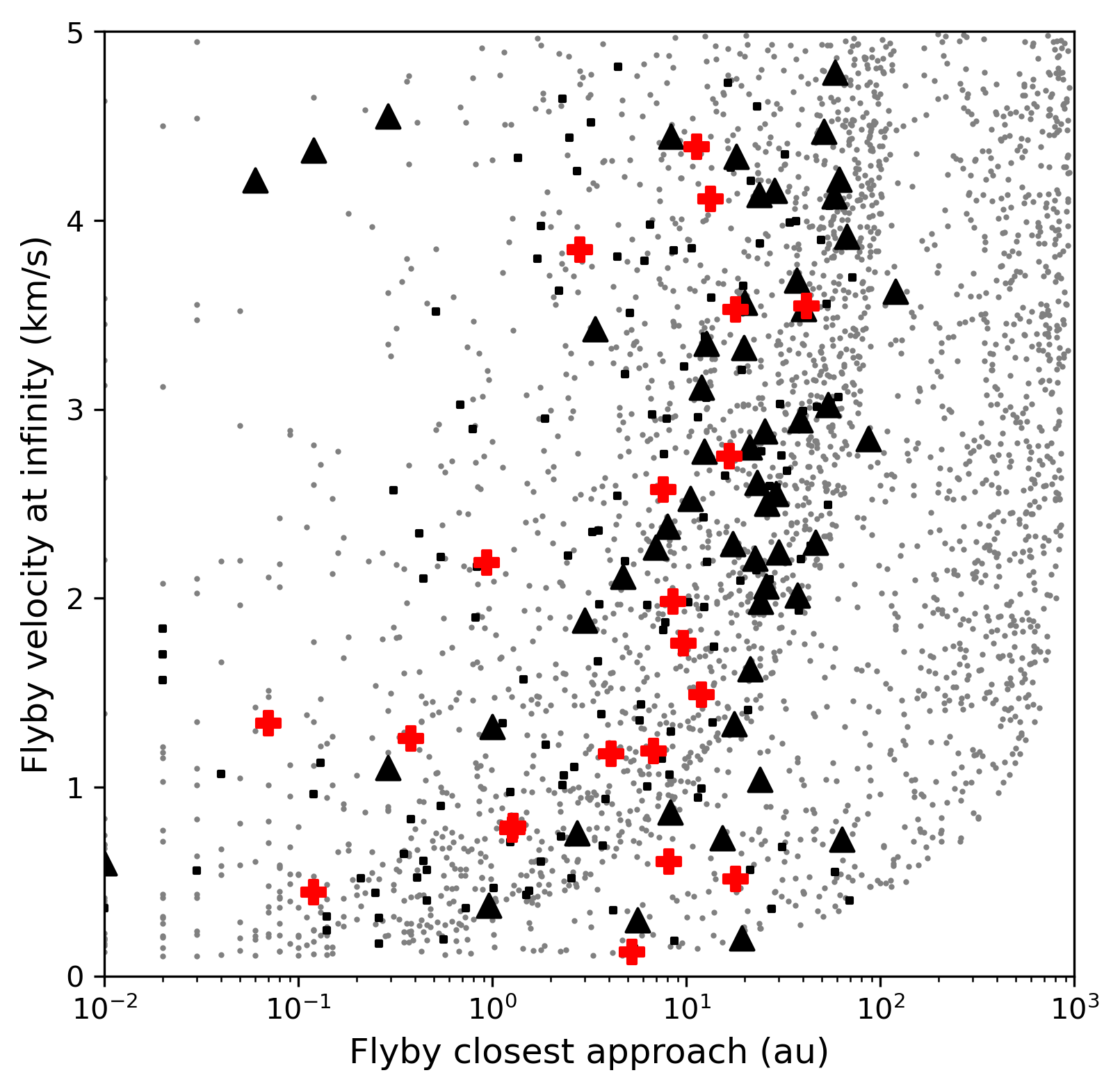}   
	\includegraphics[height=0.3\columnwidth]{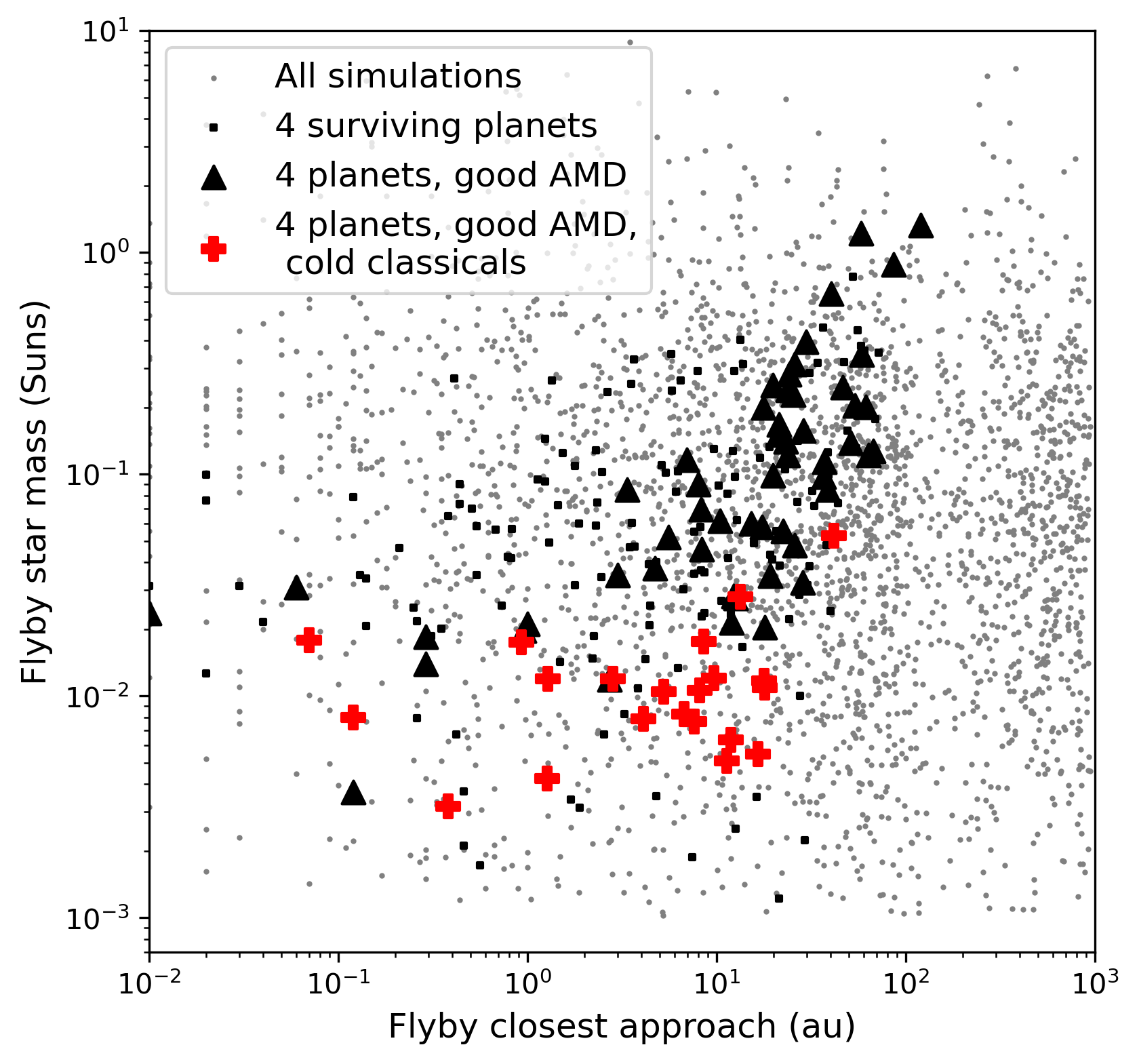}   
	\includegraphics[height=0.3\columnwidth]{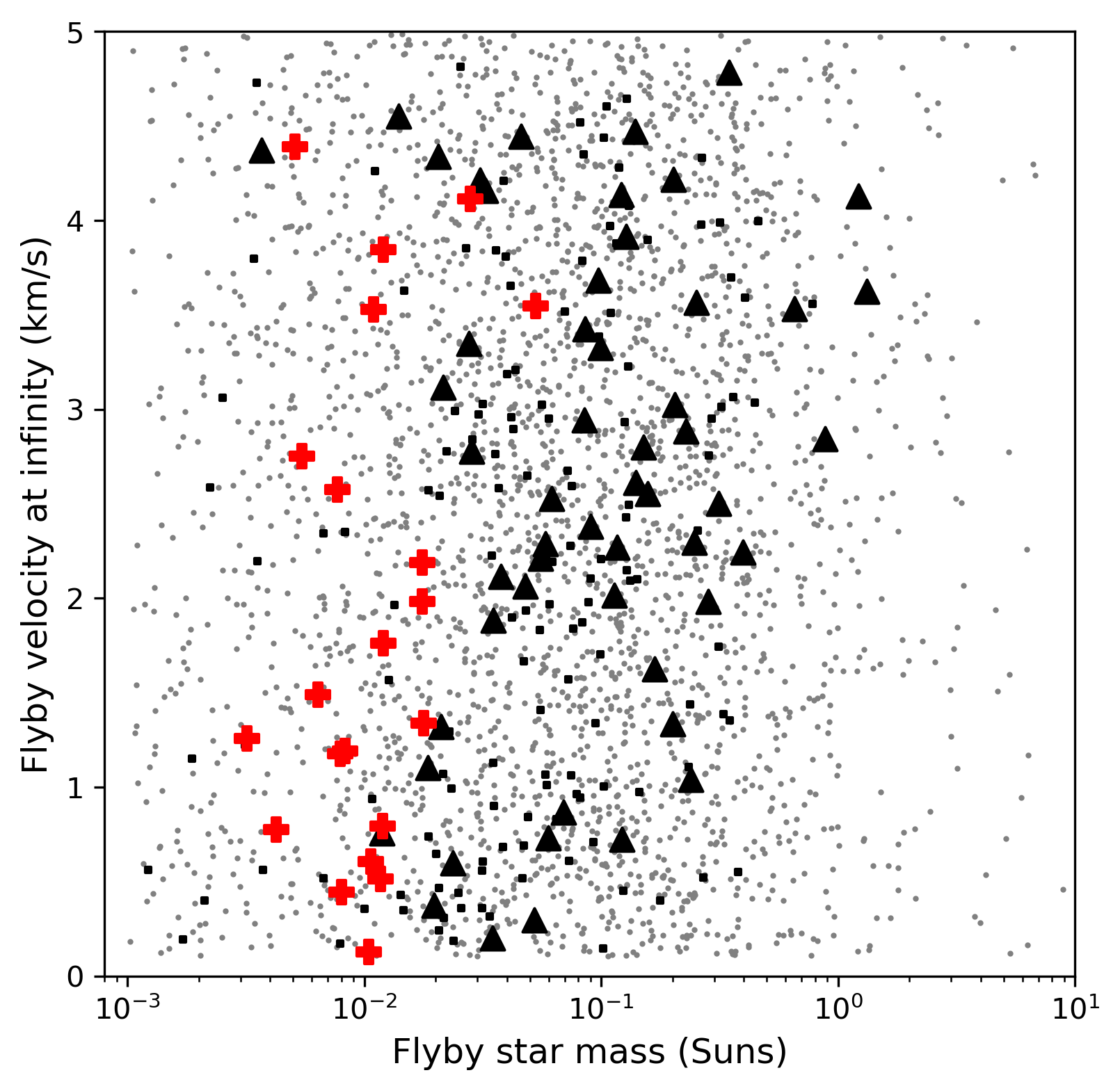}
	 \caption{Outcomes of our 3000 simulations in the parameter space of stellar flybys.  Each symbol corresponds to a single simulation.  The black filled circles are simulations in which four giant planets survived.  The large black triangles are cases where the four giant planets survived in the correct order and with an $AMD$ value within a factor of three of the present-day giant planet system.  The red crosses are systems that matched the giant planets' $AMD$ and also retained at least 40\% of their cold classical Kuiper belt particles, with a contamination rate in the hot Kuiper belt population below 40\% (see Section 3.3).}
	  
     \label{fig:outcomes}
\end{centering}
\end{figure*}

Figure~\ref{fig:outcomes} shows the broad-strokes outcomes of simulations as a function of the three flyby parameters: the flyby star (or substellar object) mass, velocity at infinity $\vinf$, and closest approach distance.  Simulations in which no dynamical instability was triggered were preferentially found in systems in which the flyby was fast (high $\vinf$) and distant (large $b$) and the flyby star was low in mass.\footnote{The strength of a stellar flyby can be measured in different ways.  The impulse that a flyby imparts on the Sun is simply $\frac{2 G M_\star}{\vinf b}$, where $b$ is the impact parameter, $M_\star$ is the mass of the flyby star or substellar object and $G$ is the gravitational constant. The impulse gradient $\frac{2 G M_\star}{\vinf b^2}$, which measures the radial gradient in acceleration felt by planets orbiting the Sun, has also been shown to be an important parameter governing the outcome of flybys for planetary systems~\citep{raymond24,kaib24a}.}
Dynamical instability was only triggered in simulations with closest approach distances less than $\sim$100 au.  This holds among simulations with large impact parameters, which must have very low $\vinf$ in order to have strong enough gravitational focusing to be come within 100 au. The simulation with the largest impact parameter that matched the giant planets' orbits had $b = 744$~au and $\vinf = 0.13$~km/s, for a closest approach distance of 5.3 au.

\subsection{Matching the giant planets}

We now turn our focus to systems that have the potential to match the Solar System.  The most important factor is to have the right number of surviving planets (and the right ones).  From our 3000 simulations, 75 had four surviving giant planets, including Jupiter and Saturn (in the correct order) and two ice giants, with the outermost ice giant's semimajor axis no larger than 35 au, and $AMD$ values within a factor of three of the present-day Solar System -- that is, between 1/3 and 3 times the Solar System value. 

\begin{figure*}
	\includegraphics[width=0.49\columnwidth]{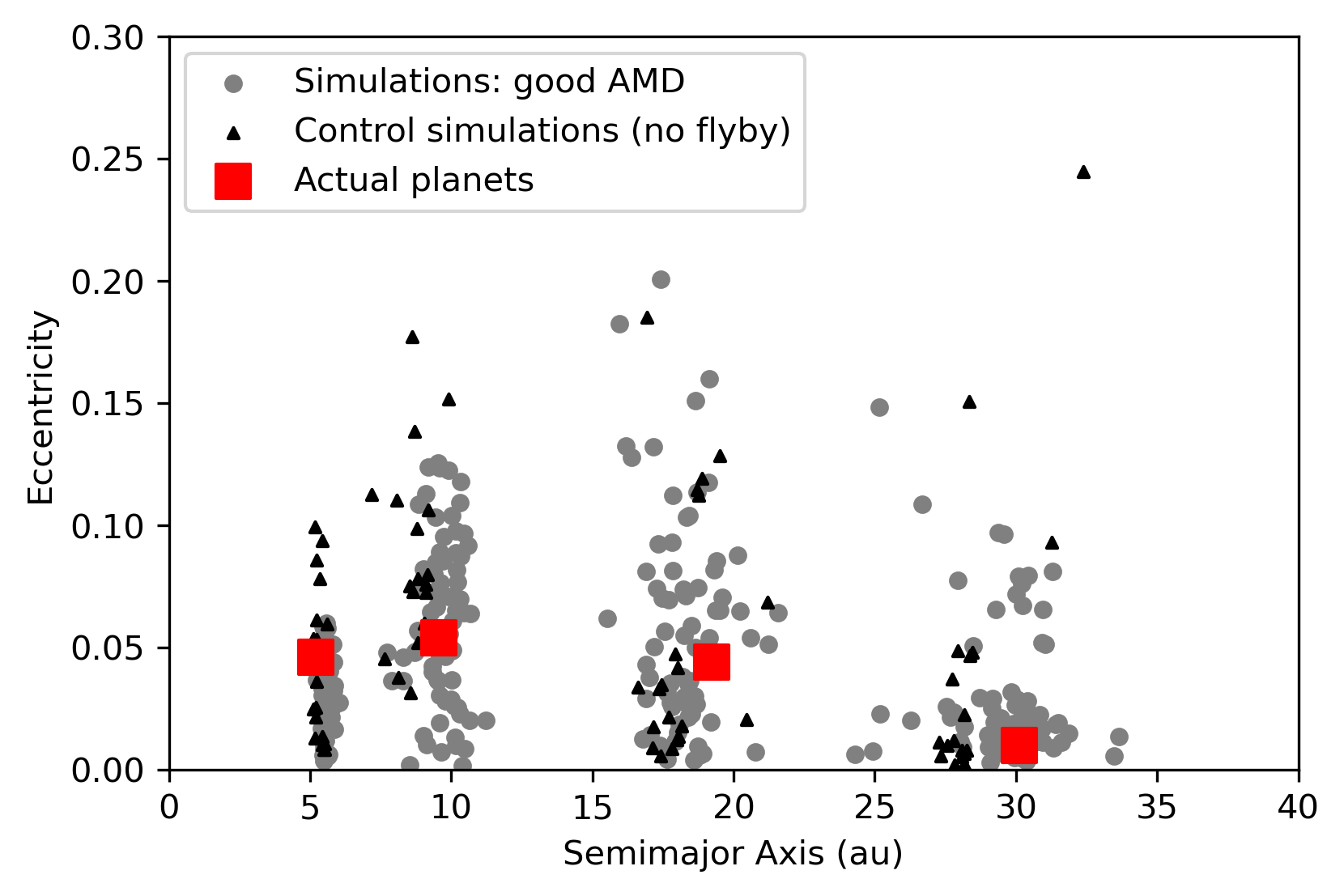}
	\includegraphics[width=0.49\columnwidth]{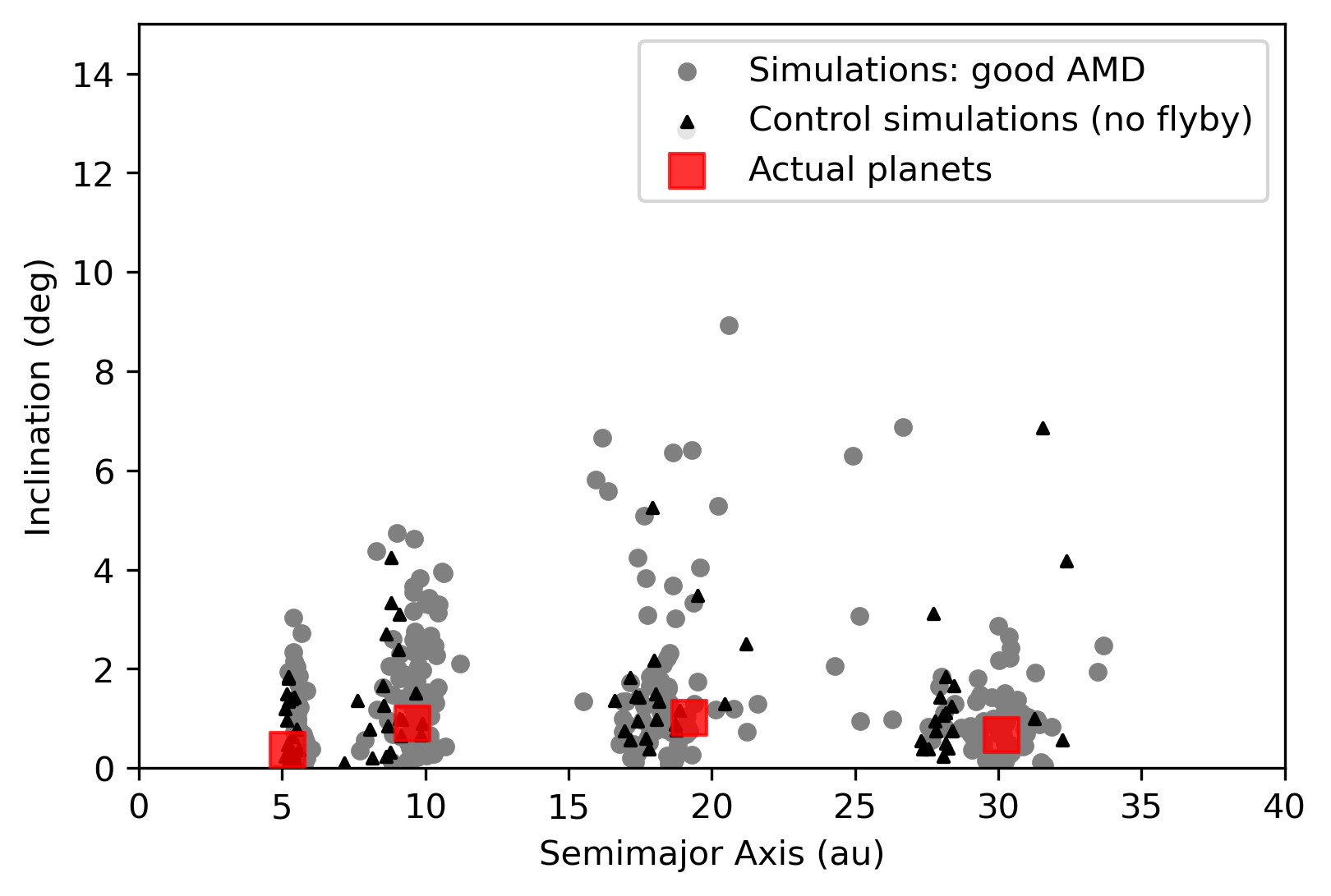}   
 \caption{Final orbits of the giant planets in 75 simulations with four surviving giant planets and $AMD$ values within a factor of three of the present-day system, as compared with the actual (time-averaged) orbits of the giant planets~\citep[taken from][]{nesvorny12}. The blue symbols are a sample from \cite{kaib16} in which the giant planets started in the same orbital configuration and four planets survived, with no flyby. }
    \label{fig:finalorbits}
\end{figure*}

Figure~\ref{fig:finalorbits} shows the orbits of the giant planets in the 75 simulations with four surviving giant planets and $AMD$ values within a factor of three of the present-day system.  Their distribution matches the present-day orbits of the giant planets roughly as well as other studies of the giant planet instability~\citep[e.g.][]{nesvorny12,clement21a,liu22}.  Fig.~\ref{fig:finalorbits} also includes the outcome of a control sample from \cite{kaib16} in which the giant planets started in the same multi-resonant orbital configuration and went unstable without a flyby, albeit with a slightly different planetesimal disk profile, extending from 21.4 to 30 au (also containing $20 \mearth$) and also following a $1/r$ surface density profile. Our flyby simulations do a better job in matching the orbits of the ice giants for this reason, although our simulations tend to place Uranus somewhat closer to the Sun than its present-day orbit. 

These simulations tend to modestly under-excite Jupiter's eccentricity. Jupiter's final eccentricity is typically 0.02-0.03 rather than its current (10 Myr-averaged) value of 0.046.  Jupiter's eccentricity was higher than 0.046 in only 11 of 75 simulations (14.7\%) that matched the giant planets' orbits. This is a known problem with the giant instability model: the degree to which the planets' orbits spread out during the instability correlates with Jupiter's final eccentricity, and simulations that match the radial extent of the Solar System tend to under-excite Jupiter's eccentricity~\citep{nesvorny12}.  The solution to this problem appears to be that Jupiter's eccentricity was already excited to close to its current value before the instability (Clement et al 2021a, 2021b), which is a natural result of interactions with the gaseous disk for certain parameters~\citep{pierens14,ragusa18,wafflard25}.  

\begin{figure*}
	\includegraphics[width=0.32\columnwidth]{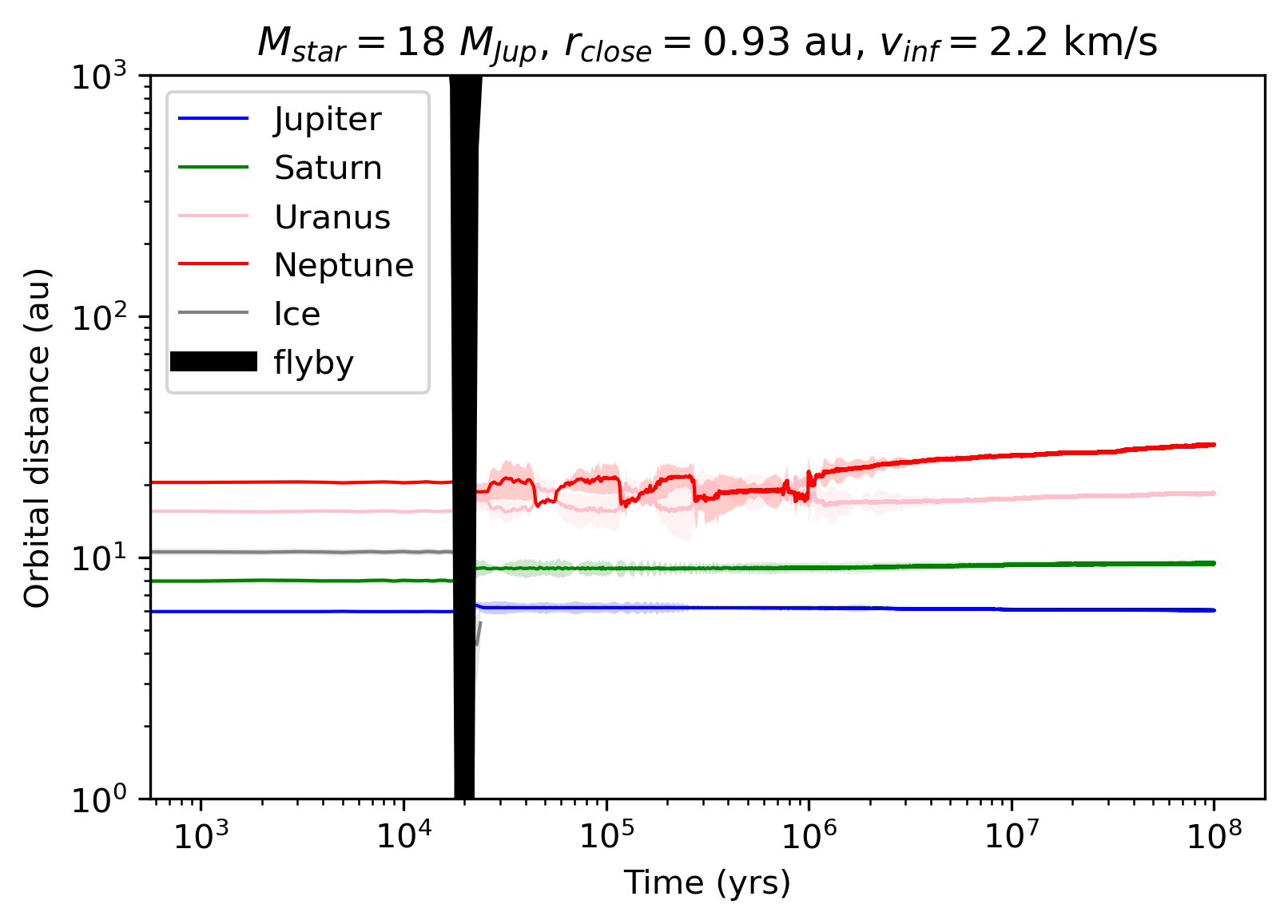}
	\includegraphics[width=0.32\columnwidth]{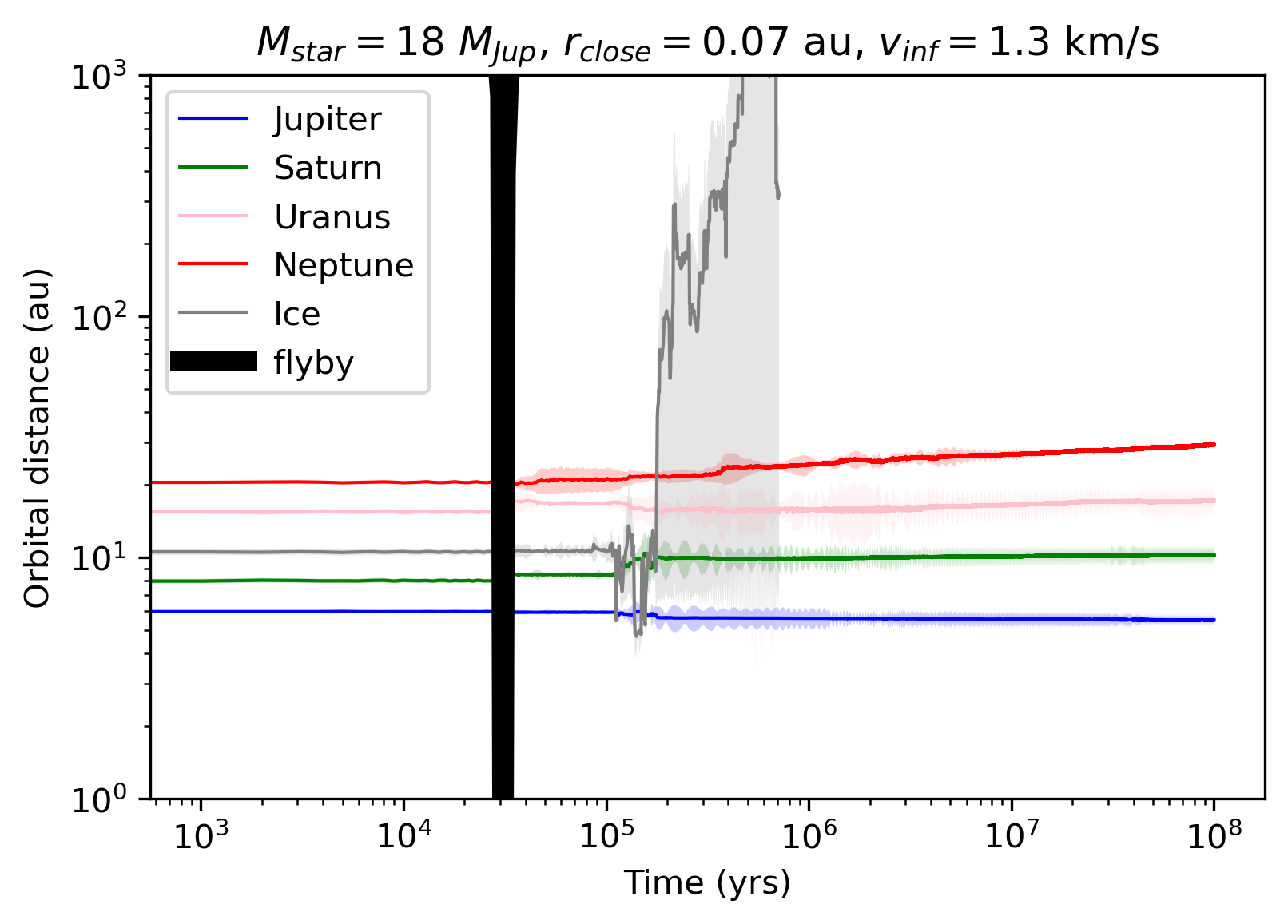}
	\includegraphics[width=0.32\columnwidth]{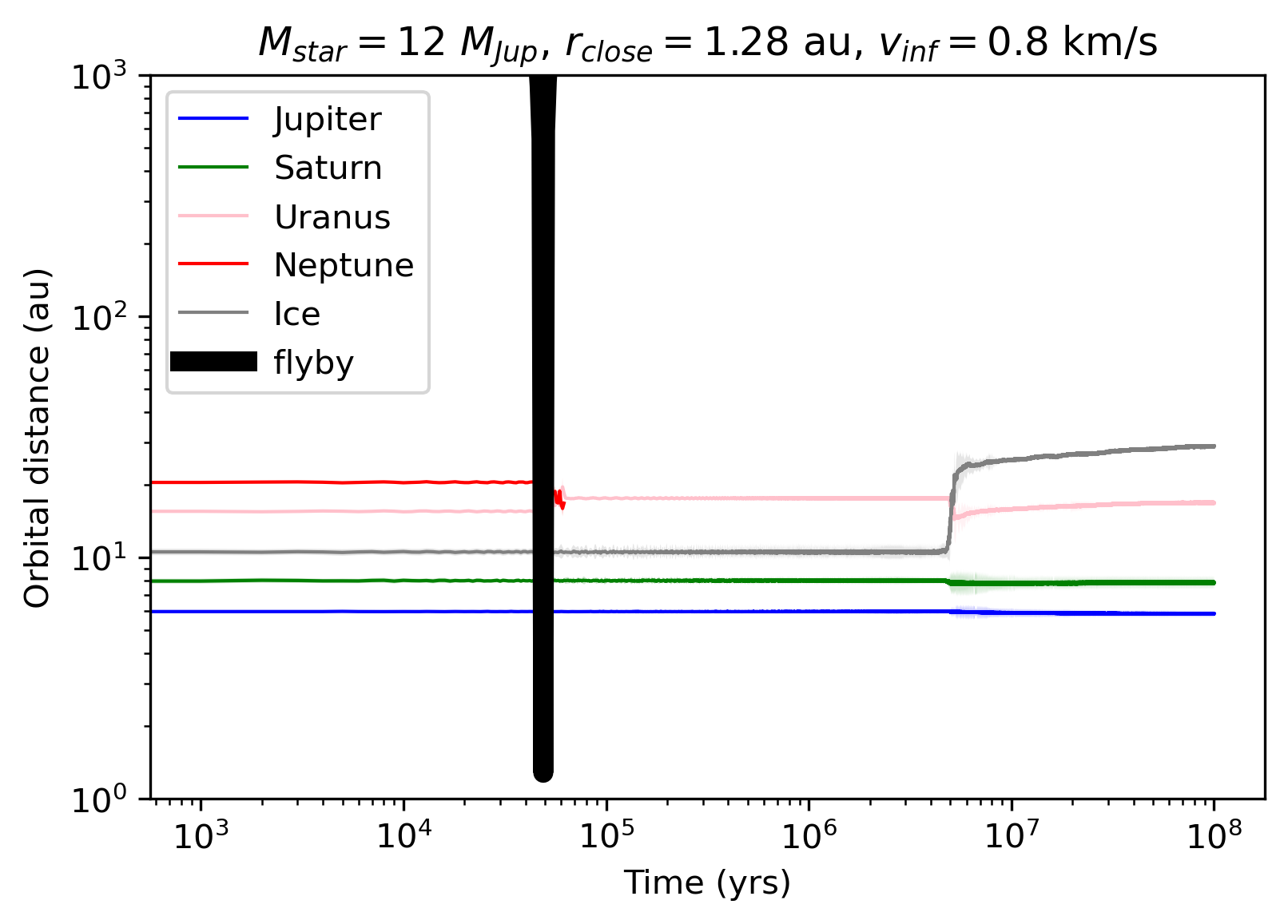}\\
	\includegraphics[width=0.32\columnwidth]{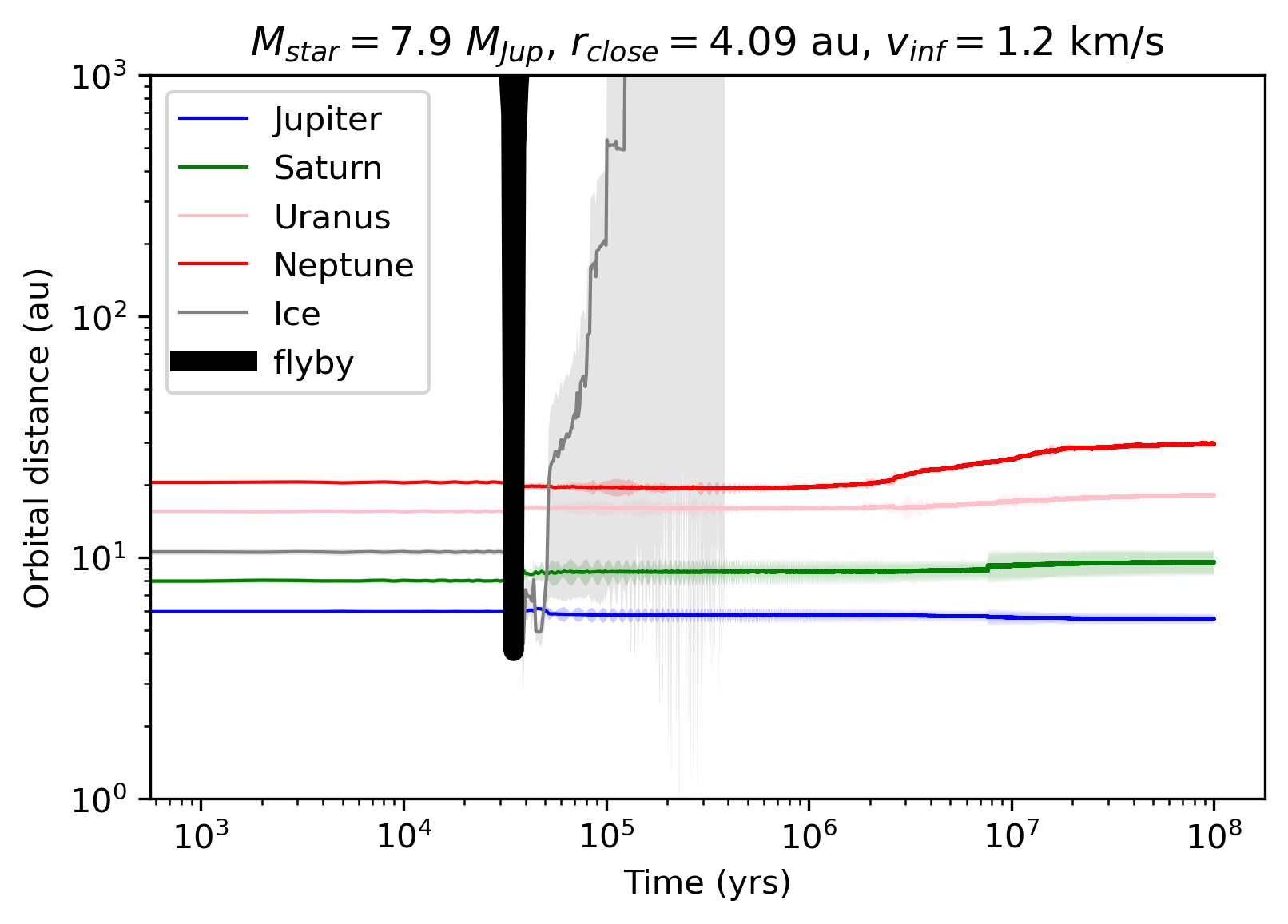}
	\includegraphics[width=0.32\columnwidth]{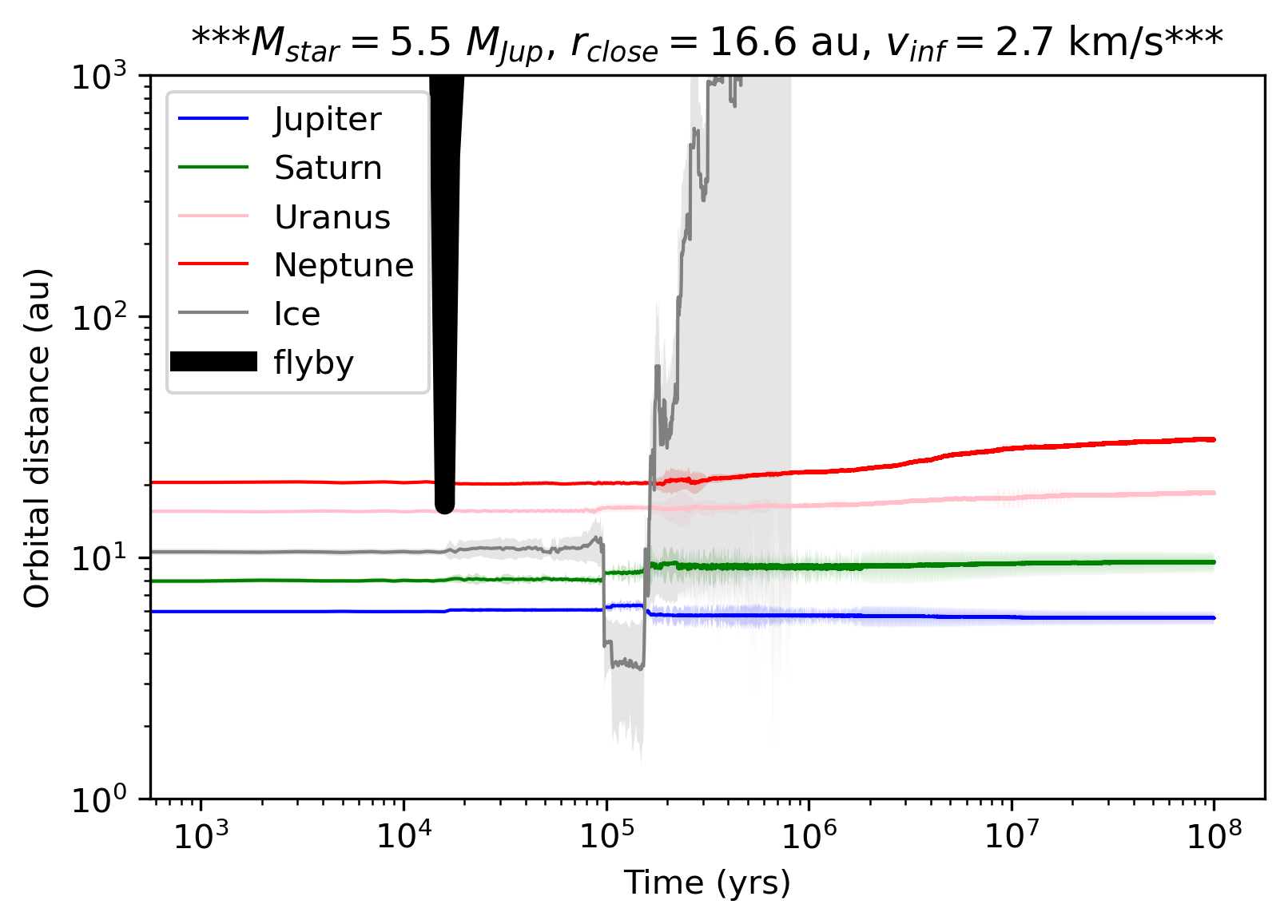}
	\includegraphics[width=0.32\columnwidth]{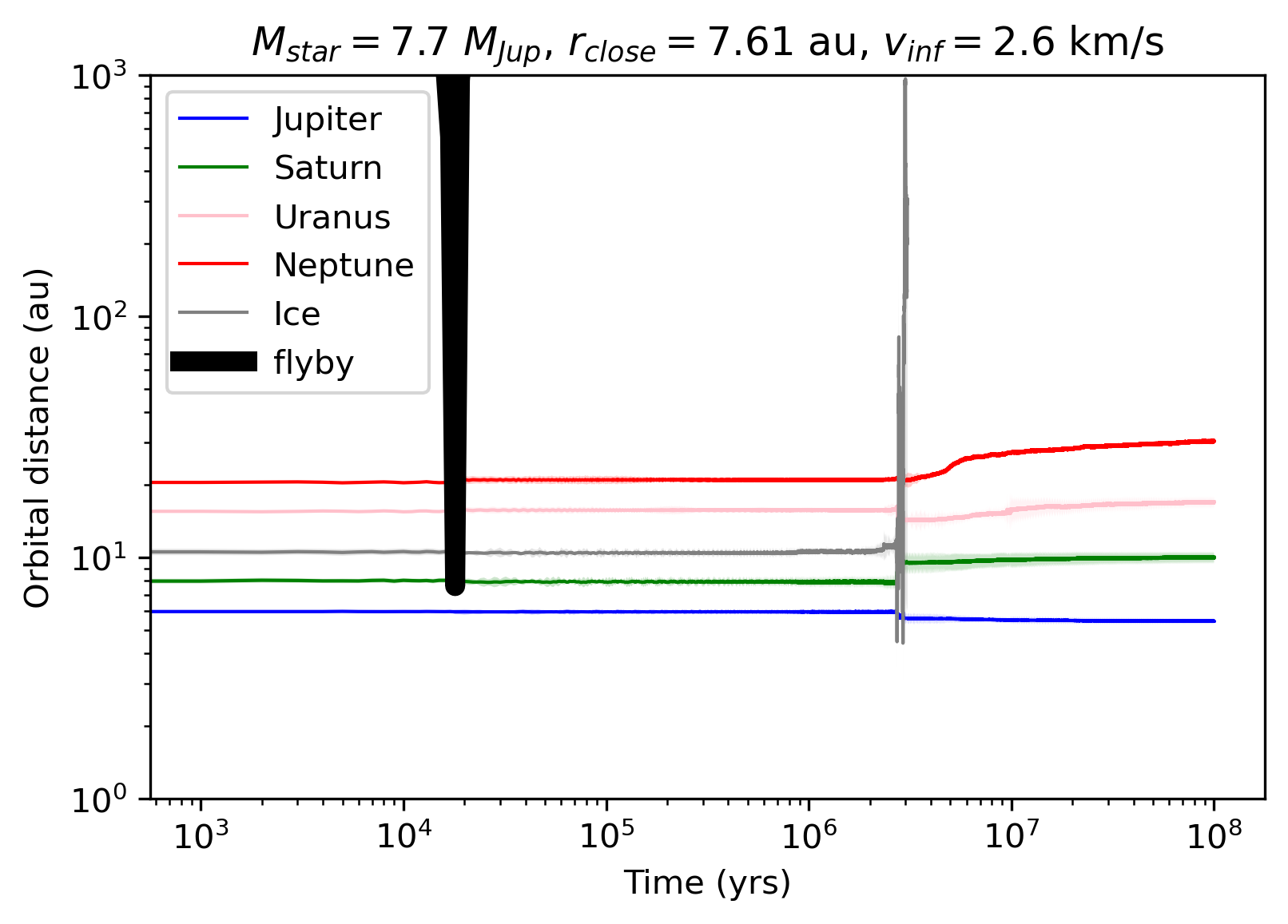}\\
	\includegraphics[width=0.32\columnwidth]{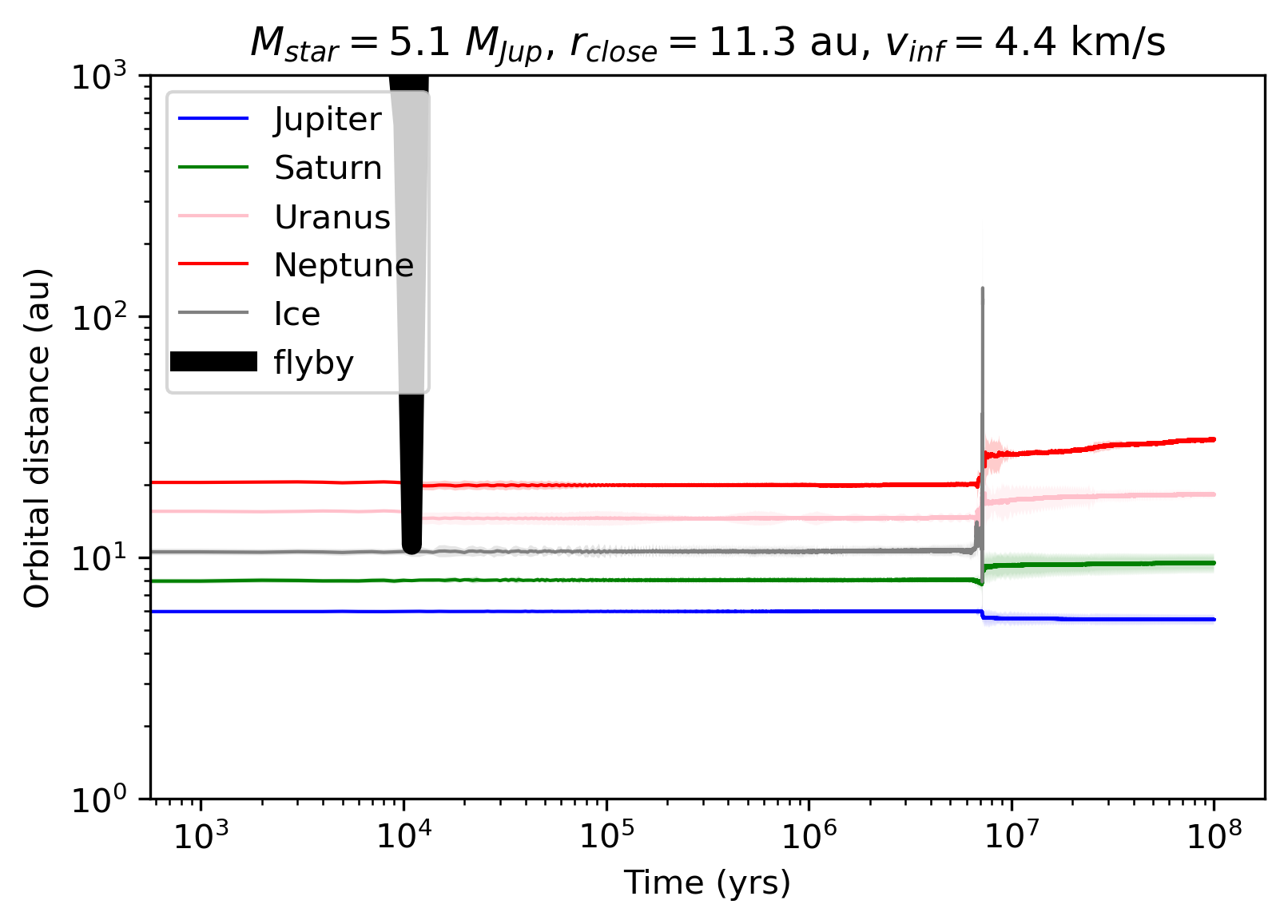}
	\includegraphics[width=0.32\columnwidth]{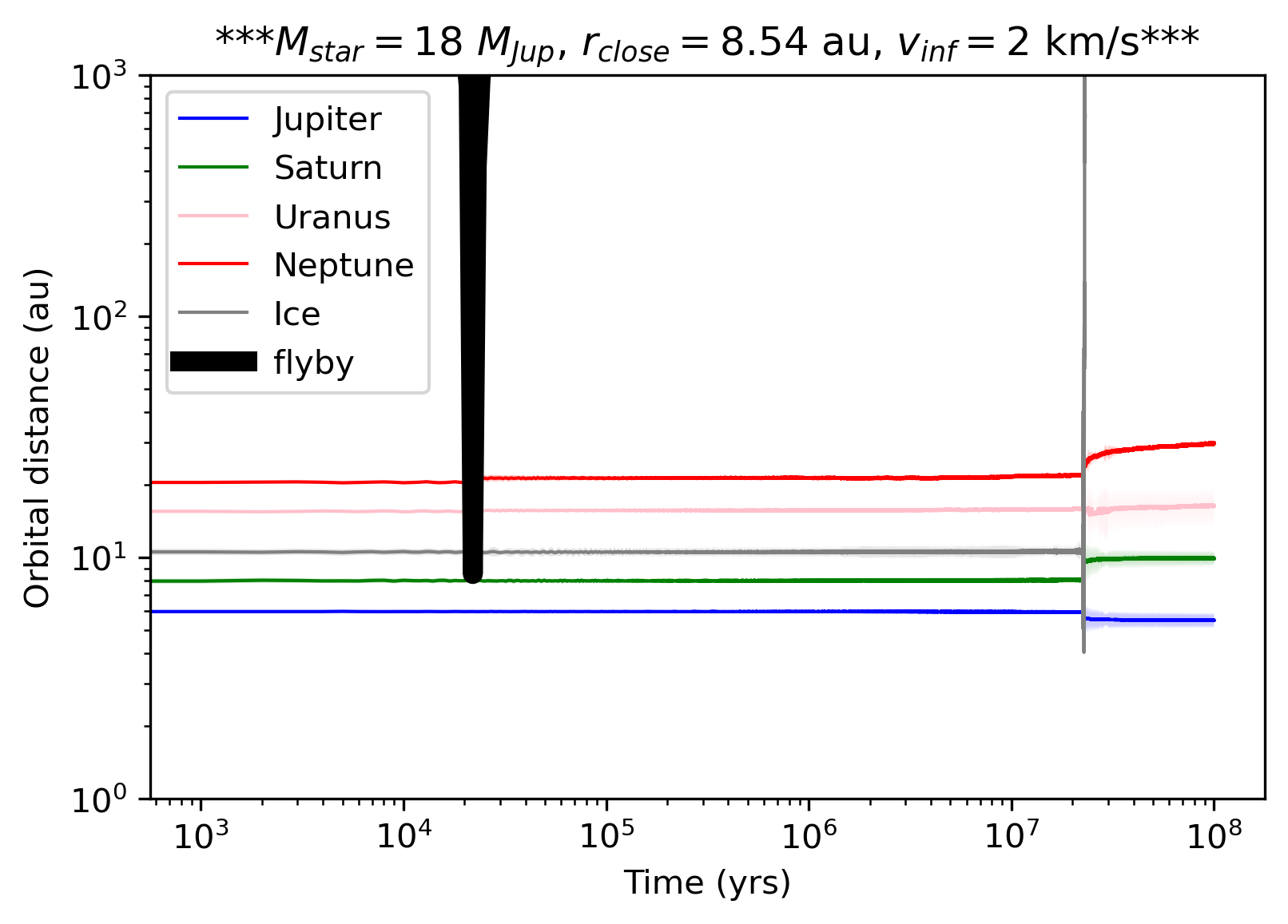} 
	\includegraphics[width=0.32\columnwidth]{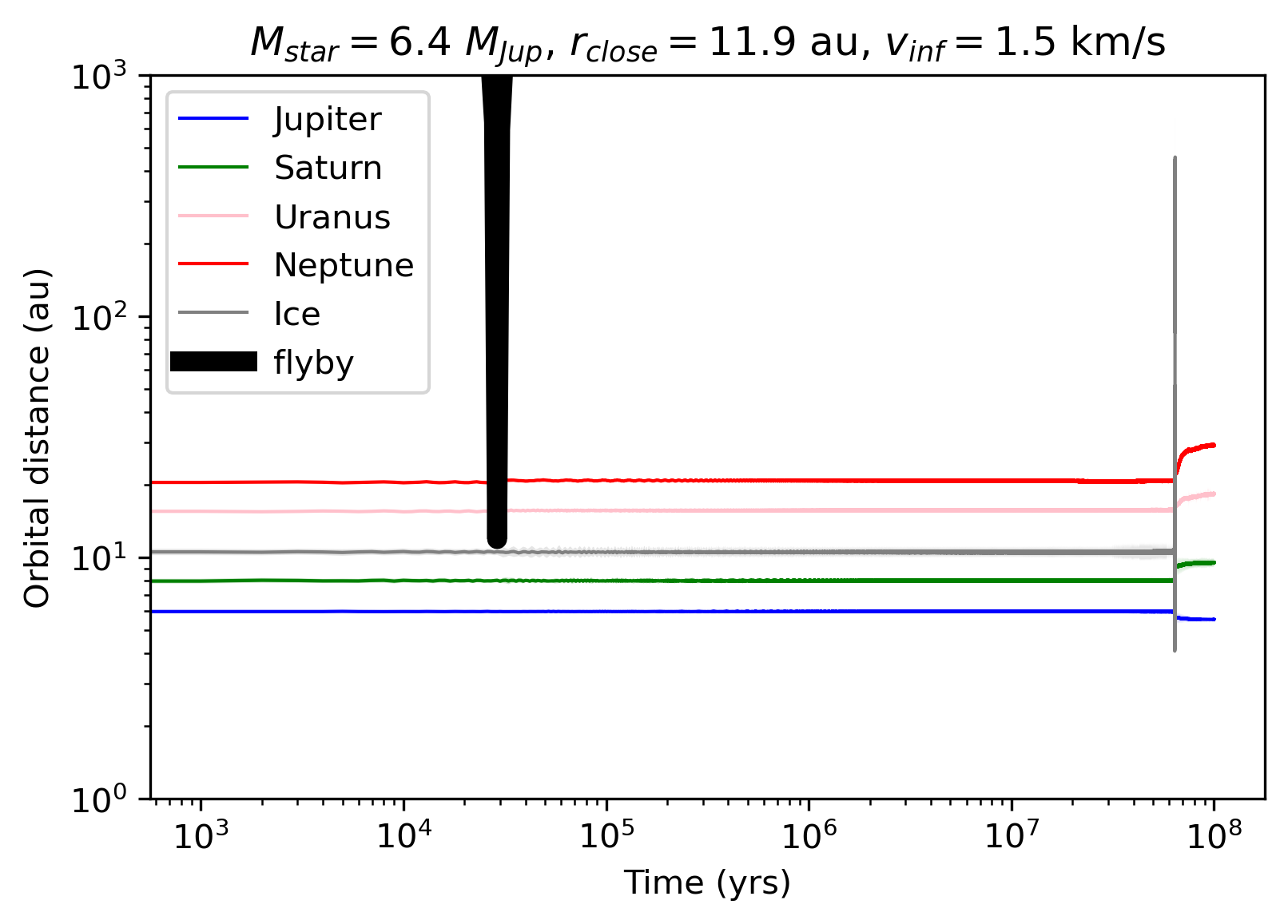}\\
		 \caption{Evolution of nine simulations that produced a good match to the giant planets' orbits while also retaining at least 40\% of their cold classical Kuiper belt particles.  Each panel shows the evolution of the orbital semimajor axis of each giant planet, with the thickness of each curve representing the eccentricity (via the perihelion to aphelion range), labeled with the flyby parameters (the closest approach of the flyby star is shown, which is a combination of the impact parameter and velocity at infinity).  These examples are ordered by the time of the onset of instability.  The two panels with stars (top center and bottom center) are featured in Fig.~\ref{fig:examples}. }
     \label{fig:evol}
\end{figure*}

\begin{figure*}
	\includegraphics[width=0.49\columnwidth]{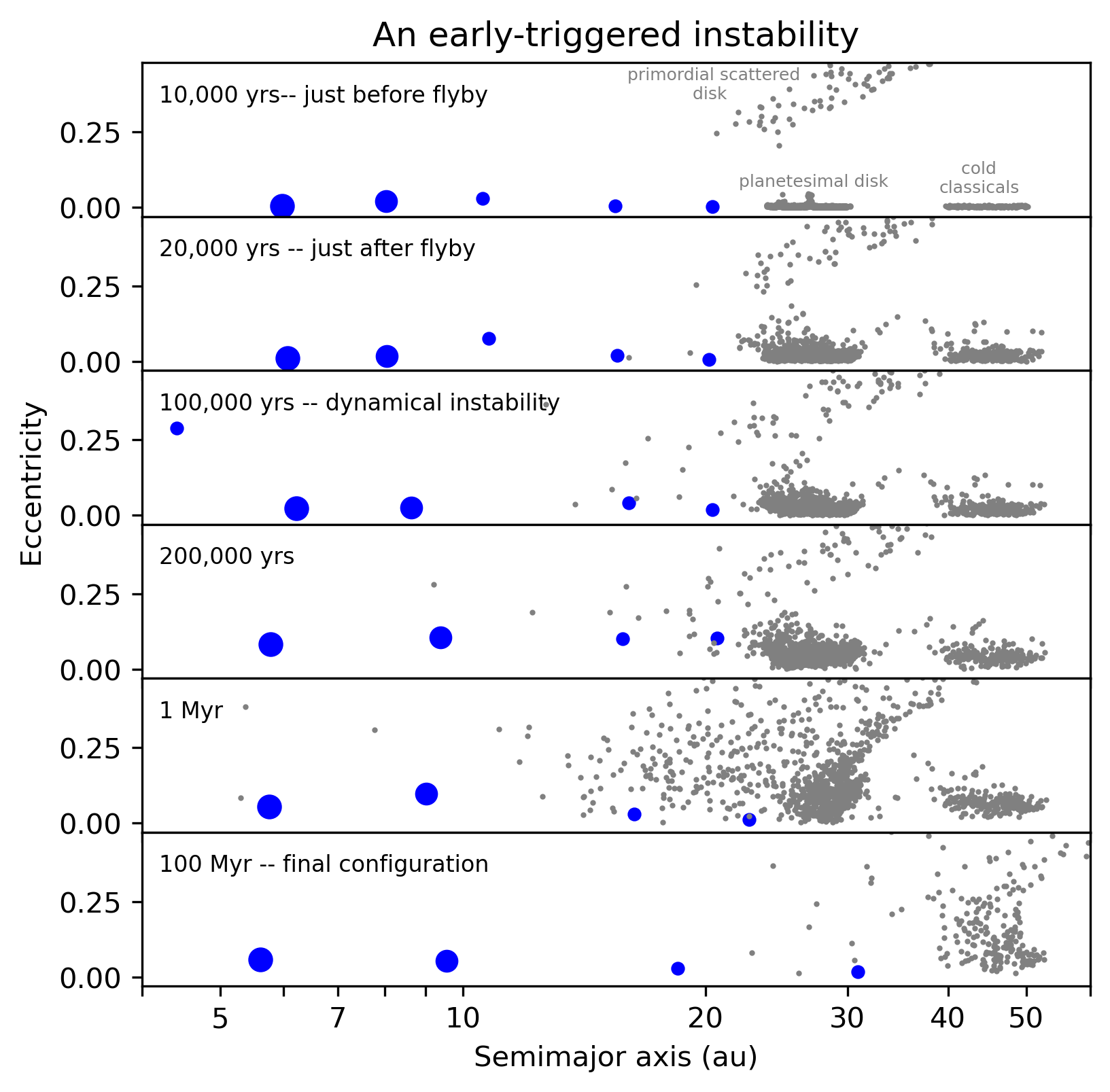}
	\includegraphics[width=0.49\columnwidth]{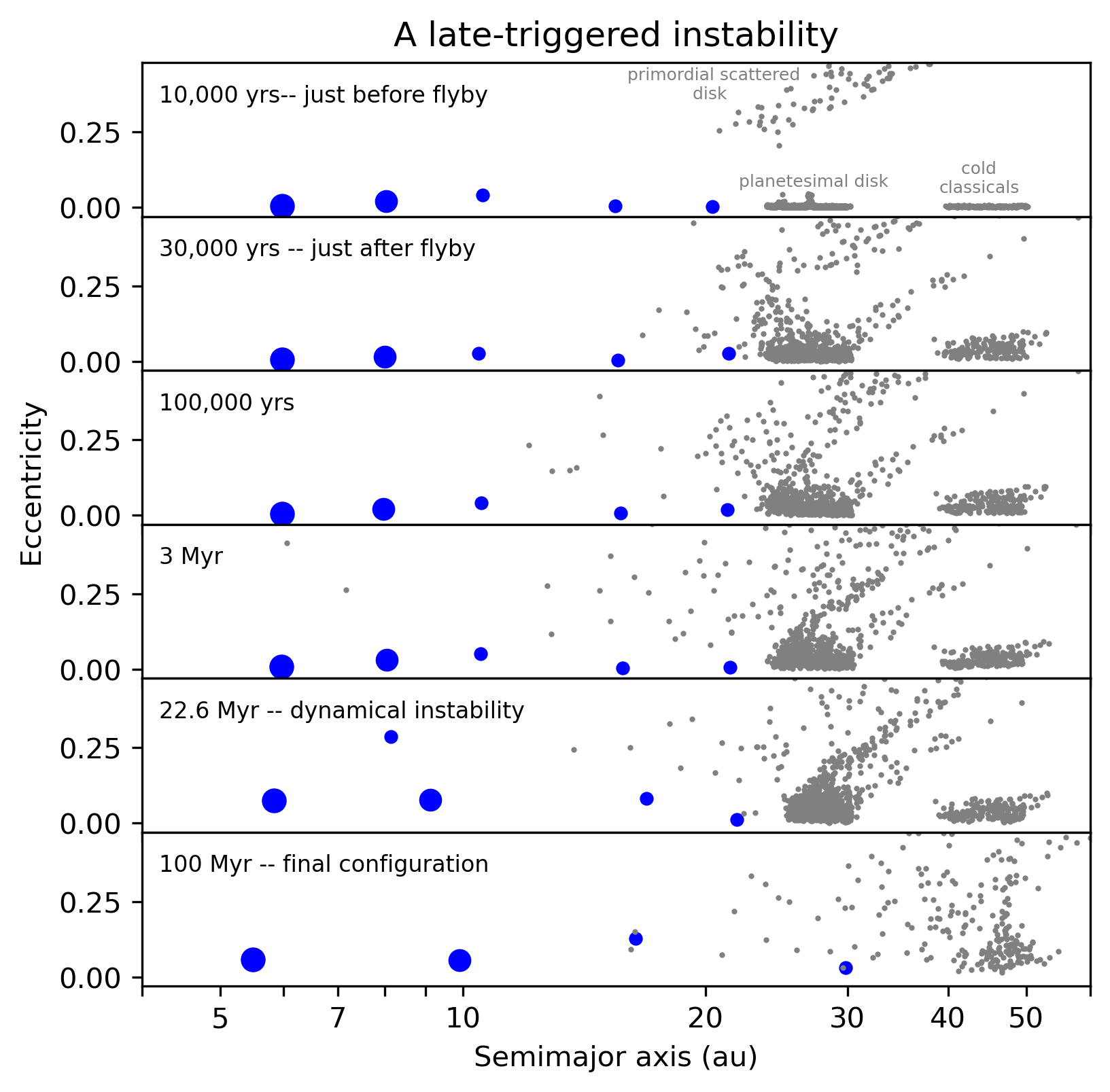}   
 \caption{Snapshots in the evolution of two simulations that both provided a good match to the present-day Solar System yet had very different instability times.  The early instability (left panel) was triggered less than $10^5$ years after the flyby of a $5.4 \mjup$ free-floating planet ($\vinf =2.7$~km/s, impact parameter of 64.5 au, closest approach of 16.6 au).  In contrast, the late instability (right panel) was only triggered 22.5 Myr after the flyby of a $17.5 \mjup$ brown dwarf ($\vinf =2$~km/s, impact parameter of 63.2 au, closest approach of 8.5 au).  The evolution of the giant planets in these systems is also shown in Fig.~\ref{fig:evol} (top center and bottom center panels). }
    \label{fig:examples}
\end{figure*}

Simulations that matched the giant planets' orbits had a broad range in instability timescales.  Some systems went unstable almost immediately following the flyby, whereas in other systems the instability was delayed by tens of Myr.  The latest instability time was 77 Myr, pushing up against the 100 Myr duration of our simulations.  Figure~\ref{fig:evol} shows the evolution of the planets' orbits in nine of these simulations (each of which also matched the Kuiper belt; see discussion below), illustrating the diversity of outcomes and instability timescales.  

Figure~\ref{fig:examples} shows the full evolution of two simulations that matched the giant planets' orbits: one in which the instability was triggered rapidly and one in which the onset of instability was delayed by more than 20 Myr.  In the early instability simulation, the flyby itself induced a strong enough excitation of the innermost ice giant's eccentricity to rapidly push it to be in dynamical contact with Saturn, effectively triggering instability.  This is the usual pathway to a flyby-driven early instability, although to match the actual Solar System the flyby perturbation cannot be too strong. In the late-onset simulation from Fig.~\ref{fig:examples}, the flyby does not noticeably perturb the orbits of the planets. Rather, the flyby excites the eccentricities of particles in the inner part of the outer planetesimal disk, putting the disk in dynamical contact with the outermost ice giant. The post-flyby evolution of the system follows the evolution of a standard planetesimal disk-triggered instability, with the giant planets' orbits slowly spreading until they reach an unstable location~\citep{tsiganis05,morby07,ribeiro20}. 

Among simulations that matched the giant planets' orbits, those that underwent an early-triggered instability tend to have undergone a somewhat closer approach from a lower-mass flyby object. If we divide the 75 cases that matched the giant planets' orbits into simulations in which the instability was triggered before vs. after 1 Myr.  The early-triggered instabilities had a median flyby mass of $29 \mjup$ and a median impact parameter of 49.85 au (with a median closest approach distance of 6.3 au).  In contrast, the late-triggered instabilities had a median flyby mass of $85 \mjup$, with a median impact parameter and close approach distance of 80.7 and 18 au, respectively.  This reinforces our previous claim that early-triggered instabilities underwent flyby-driven perturbations the planets directly, whereas it was the planetesimal disk that was mainly perturbed in the late instabilities.  It is worth noting that the final outcomes of these simulations do not show any clear differences between simulations in which the instability was triggered early or late, in terms of quantities such as the $AMD$, the orbits of the planets (including Jupiter's eccentricity), or the general (final) orbital configuration of surviving outer planetesimals.  

\subsection{Matching the Kuiper belt}

\begin{figure}
	\includegraphics[width=0.49\columnwidth]{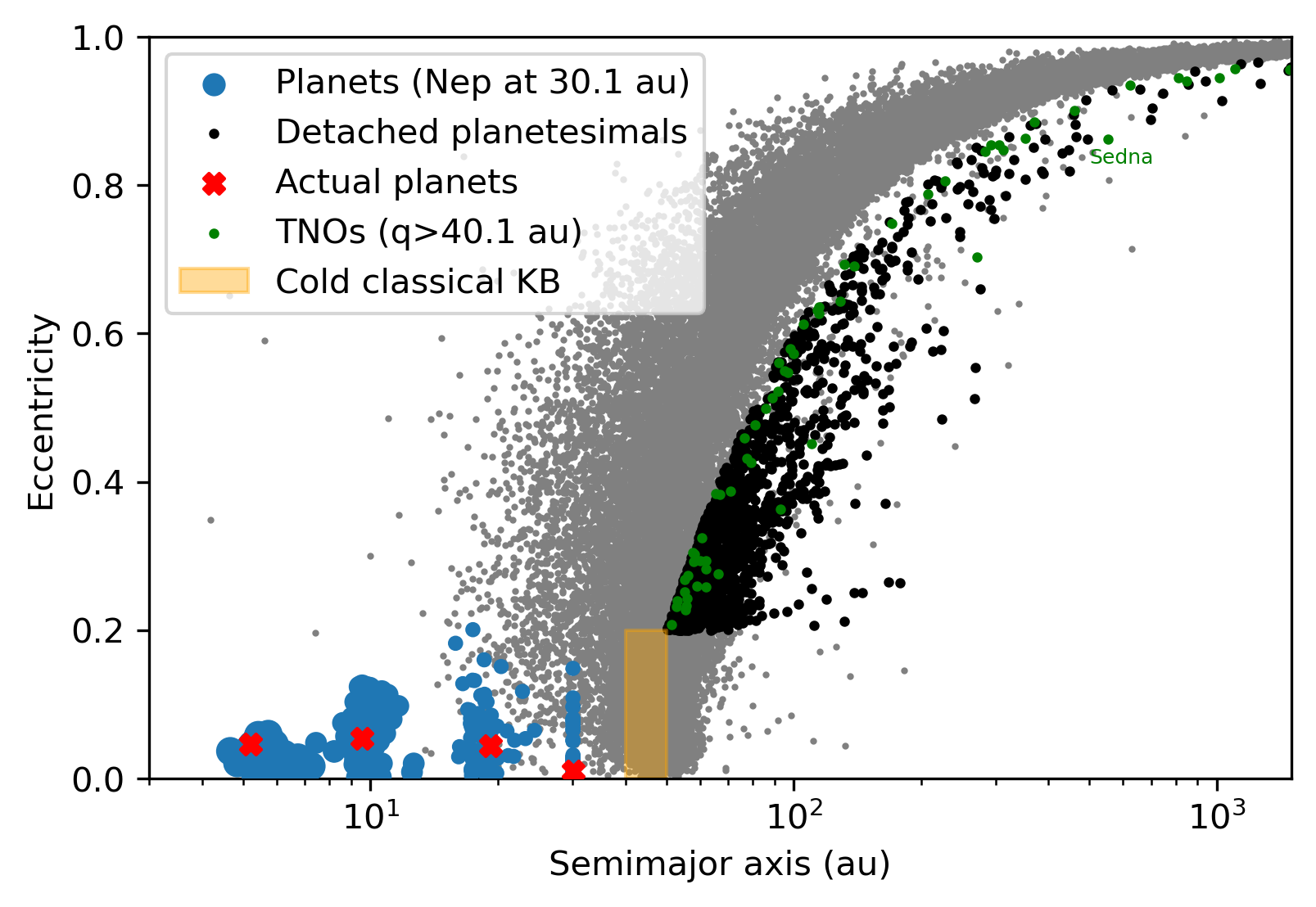}
	\includegraphics[width=0.49\columnwidth]{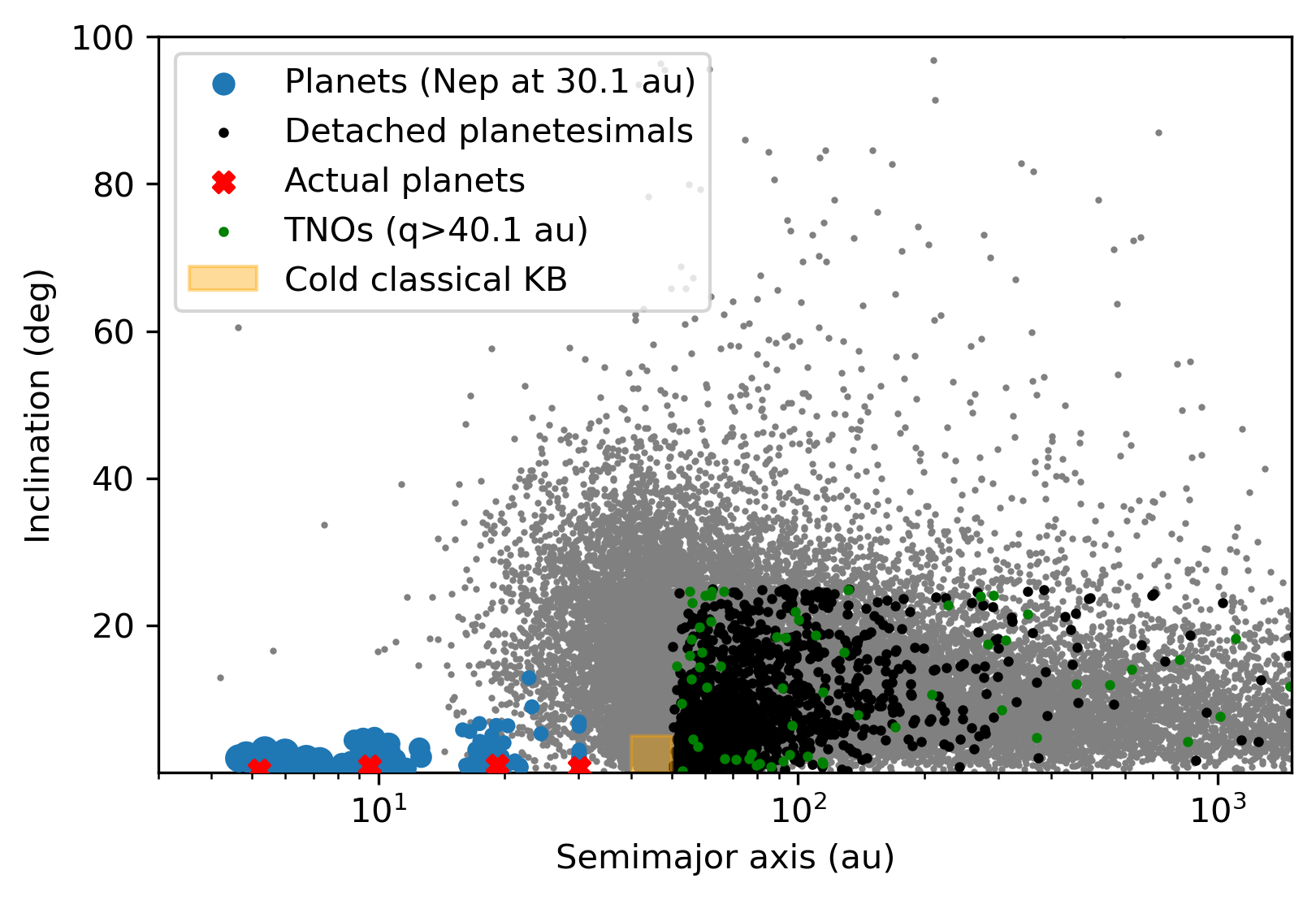}
 \caption{Final orbital configurations of the giant planets and outer planetesimals in simulations with four surviving giant planets interior to 35 au and an $AMD$ within a factor of three of the present-day Solar System value.  Here, the semimajor axes of all particles have been rescaled such that Neptune's orbital radius matches its present-day one, of 30.1 au.  Detached planetesimals are defined as those with perihelion distances larger than 40.1 au, inclinations below 25 degrees (to filter out particles in Kozai oscillation), and semimajor axes larger than 50 au. The cold classical Kuiper belt region is shaded.  Note that the clump of particles with low-eccentricity orbits past 50 au are an artefact -- they are cold classical particles in simulations in which Neptune finished interior to 30.1 au, so in this plot their orbits were widened to calibrate and compare between simulations.  In our full analysis, each simulation was considered separately. }
    \label{fig:outerSS}
\end{figure}

Figure~\ref{fig:outerSS} shows the distribution of planetesimals after 100 million years of evolution for the 75 simulations that  matched the giant planets' orbits.  Here, each system's radial extent has been normalized so that the outermost ice giant matches Neptune's present-day semimajor axis of 30.1 au.  The dominant feature is the huge population of (gray) planetesimals in the process of being ejected or implanted into the Oort cloud. Those planetesimals originated mostly in the outer planetesimal disk, with a contribution from primordial cold classical Kuiper belt and scattered disk populations.  However, they are irrelevant when evaluating the success rate in that they no longer exist on those orbits. 

We focus on matching the cold classical Kuiper belt and the population of detached TNOs.  Maintaining the observed cold classical population~\citep[see][]{brown01,petit11} is essential because its low inclination distribution puts a constraint on the strength of any flyby~\citep{batygin20}. It is likewise essential that the cold classical belt particles remain mostly confined in order to maintain the clear orbital and color separation of that population~\citep[e.g.][]{gladman21}.   In contrast, matching the detached TNO population would only be a bonus for a given flyby, as a later flyby could simply be invoked to explain those objects~\citep{nesvorny23}.  

\begin{figure}
	\includegraphics[width=0.9\columnwidth]{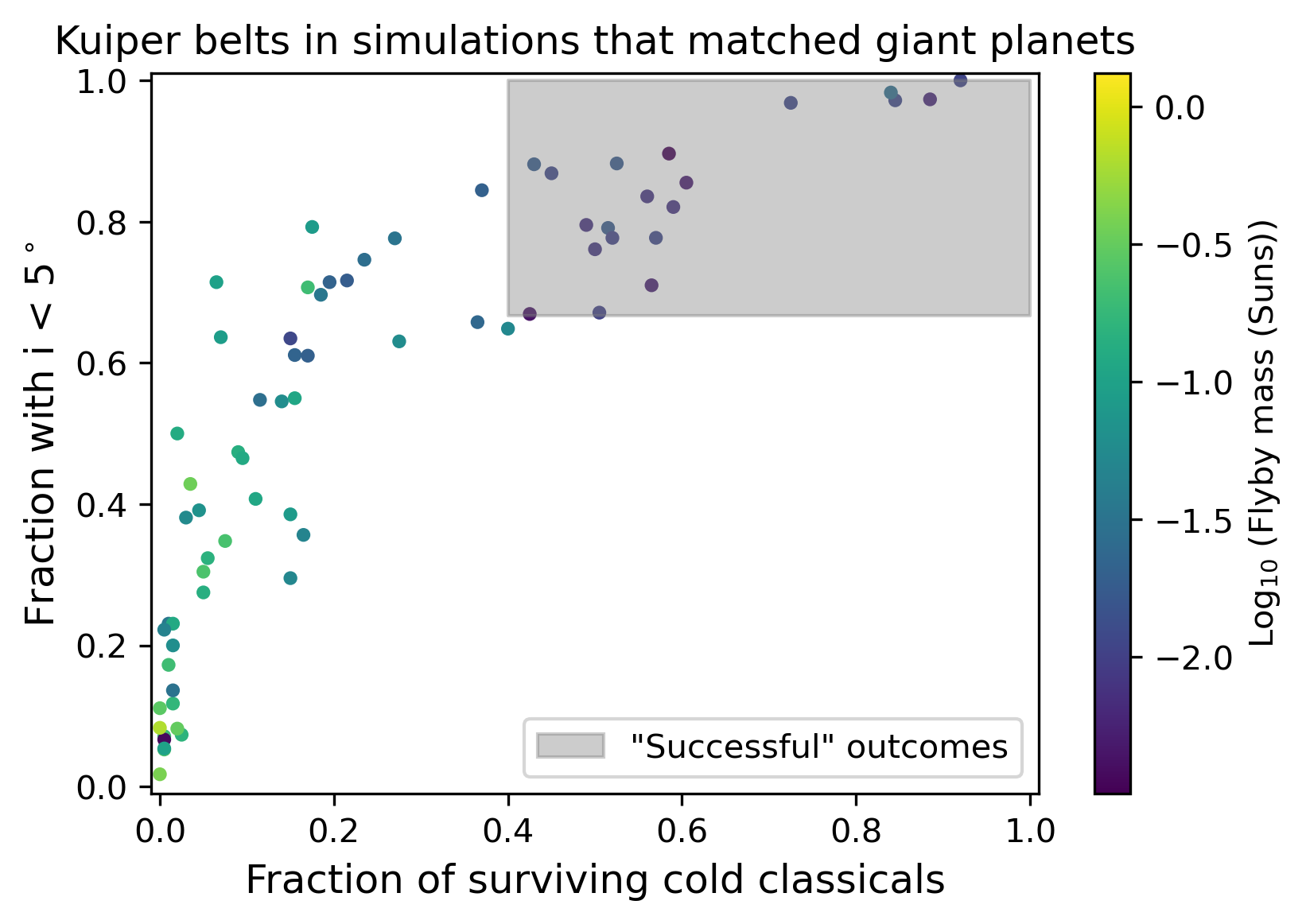}
 \caption{Cold classical Kuiper belts in the 75 simulations that matched the giant planets' orbits.  We evaluated the outcomes in terms of the fraction of the number of cold classical particles that was retained at the end of the simulation on cold orbits, and the fraction of cold classical particles that survived with inclinations smaller than $5 \deg$. The colors correspond to the logarithm of the flyby mass in each simulation. }
    \label{fig:CC}
\end{figure}

Among the 75 simulations that matched the giant planets' orbits, only a fraction retained robust cold classical Kuiper belts.  We evaluated this with two criteria.  First, we required that at least 40\% of all cold classical particles survived within the broadly-defined cold classical region in which they started, with semimajor axes between 40 and 50 au, eccentricities less than 0.2, and inclinations less than 5 degrees. Second, we required that at least two-thirds of all surviving cold classical particles (with semimajor axes between 40 and 50 au) remained on orbits with inclinations less than $5 \deg$, to respect the strong concentration at low inclinations that is observed~\citep[e.g.][]{brown01,petit11,gladman21}. In other words, we only allowed a small fraction of additional cold classical particles to survive on dynamically hot orbits. Fig.~\ref{fig:CC} shows the distribution in these quantities among simulations that matched the giant planets' orbits.  We admit that our exact cutoff values are arbitrary, chosen to balance between restricting ourselves to the simulations that best match the Solar System while keeping the success rate from dropping too low.  As such, the 20 `successful' simulations represent the best-case scenario for matching the Solar System with a flyby-driven instability.  Nonetheless, we varied these success parameters and propagated them through the rest of the analysis and found a surprisingly small effect on the conclusions of the paper (in terms of the probability of a flyby-driven instability discussed in Section 4). 

\begin{figure}
	\includegraphics[width=0.9\columnwidth]{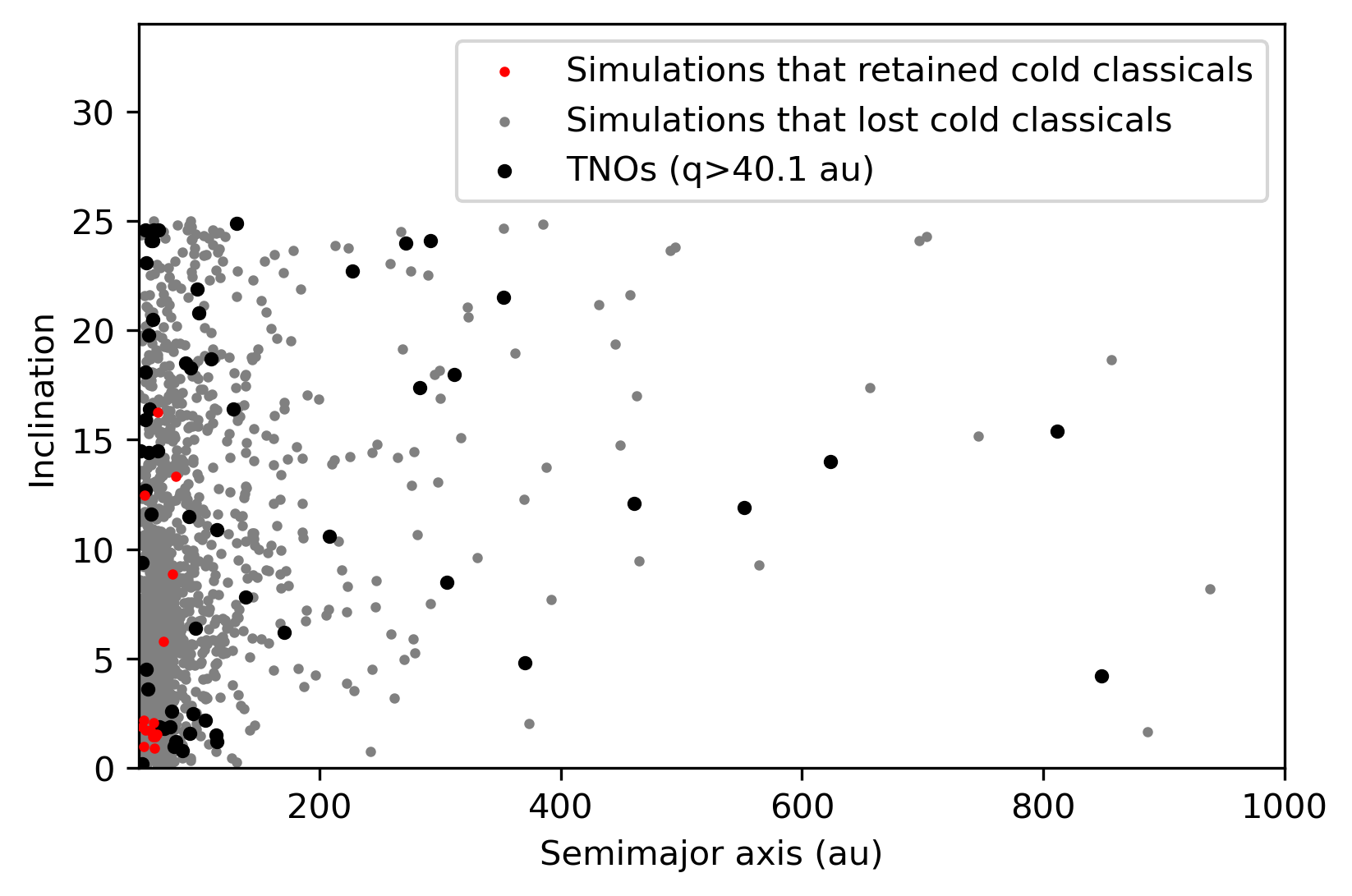}
 \caption{Orbital distribution of detached TNOs produced in simulations that matched the giant planets' orbits and retained (red) or lost (gray) the cold classical Kuiper belt population, compared with the observed distribution (black). }
    \label{fig:detached}
\end{figure}

Among simulations that matched the giant planets' orbits, we found a dichotomy between simulations that retained the cold classical Kuiper belt and those that did not. Figure~\ref{fig:detached} shows the distribution of planetesimals with detached orbits, compared with the current observational sample.  Detached orbits were defined as those with perihelion distances larger than 40.1 au, inclinations below 25 degrees (to avoid including particles in Kozai resonance with the planets), semimajor axes larger than 50 au, and eccentricities above 0.2 (after scaling each particle by the same factor that placed the Neptune analog at 30.1 au).  The broad distribution of detached objects in simulations that did not retain the cold classicals is roughly consistent with observations (although we do not attempt a careful statistical match).  In contrast, simulations that retained the cold classical Kuiper belt rarely produced detached objects with semimajor axes larger than $\sim 100$~au, and very few within the currently-observed range. The simulations that matched the giant planets but disrupted the cold classical Kuiper belt underwent modestly more violent flybys, typically with more massive objects: the median flyby for simulations that retained the cold classicals was $0.0105 \msun$ ($10.5 \mjup$) as compared with $0.094 \msun$ ($94 \mjup$) for simulations that disrupted the cold classicals.

Altogether, our sample therefore contains 20 simulations that are entirely consistent with observations, matching both the giant planets' orbits and the cold classical Kuiper belt.

\section{Putting a flyby-driven instability in the context of the Sun's birth cluster}

We now evaluate the likelihood that a flyby triggered the giant planet instability.  First, we determine the fraction of simulations that match the Solar System as a function of the flyby parameters.  Then, we build a simple model to generate flybys in a Monte Carlo fashion as a function of the configuration of the Sun's stellar birth cluster, and use this to infer the cluster properties that are consistent with a flyby-driven instability.

\subsection{Success rate as a function of flyby parameters}

\begin{figure*}
\begin{centering}
	\includegraphics[width=0.48\columnwidth]{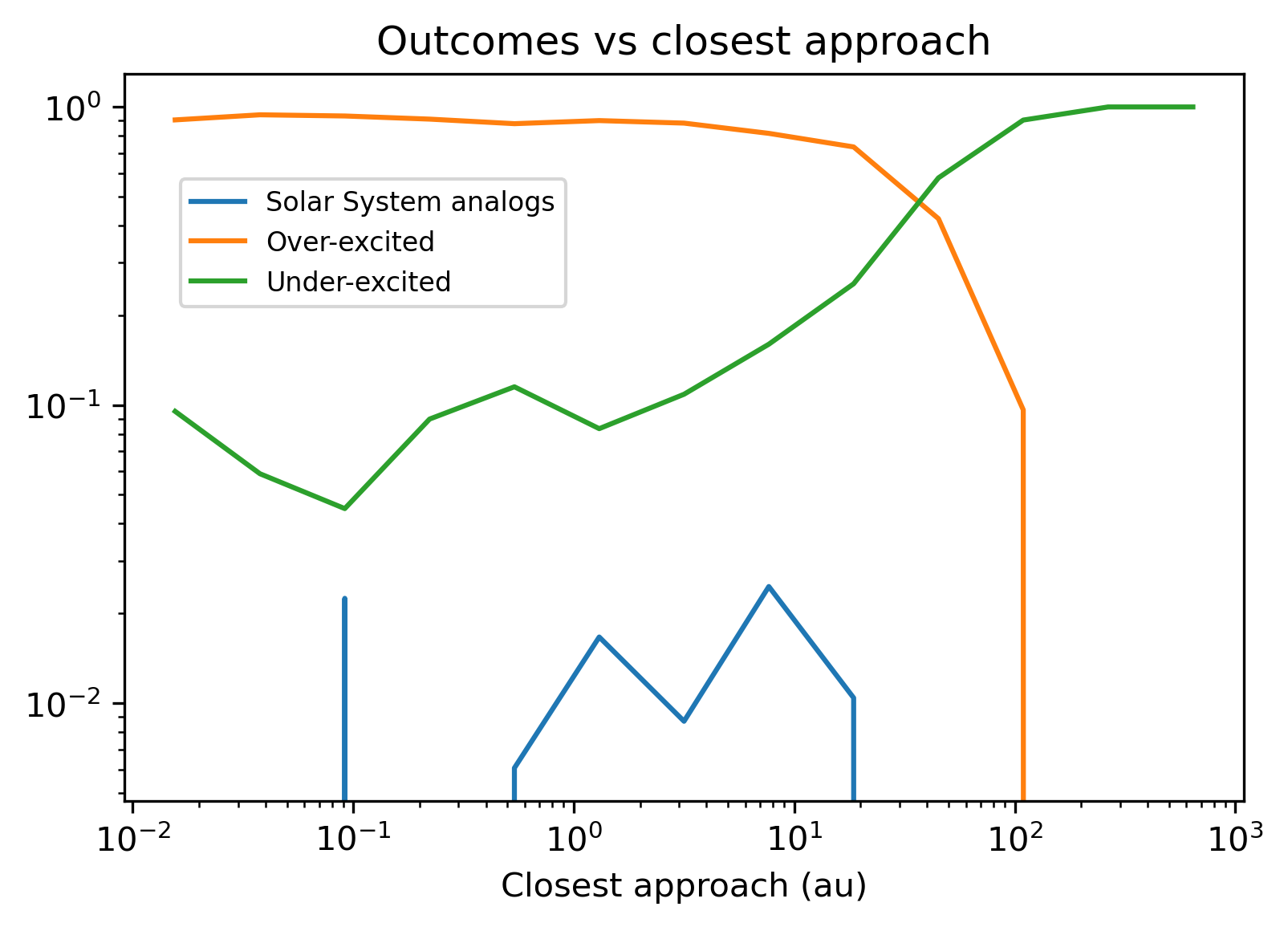}
	\includegraphics[width=0.48\columnwidth]{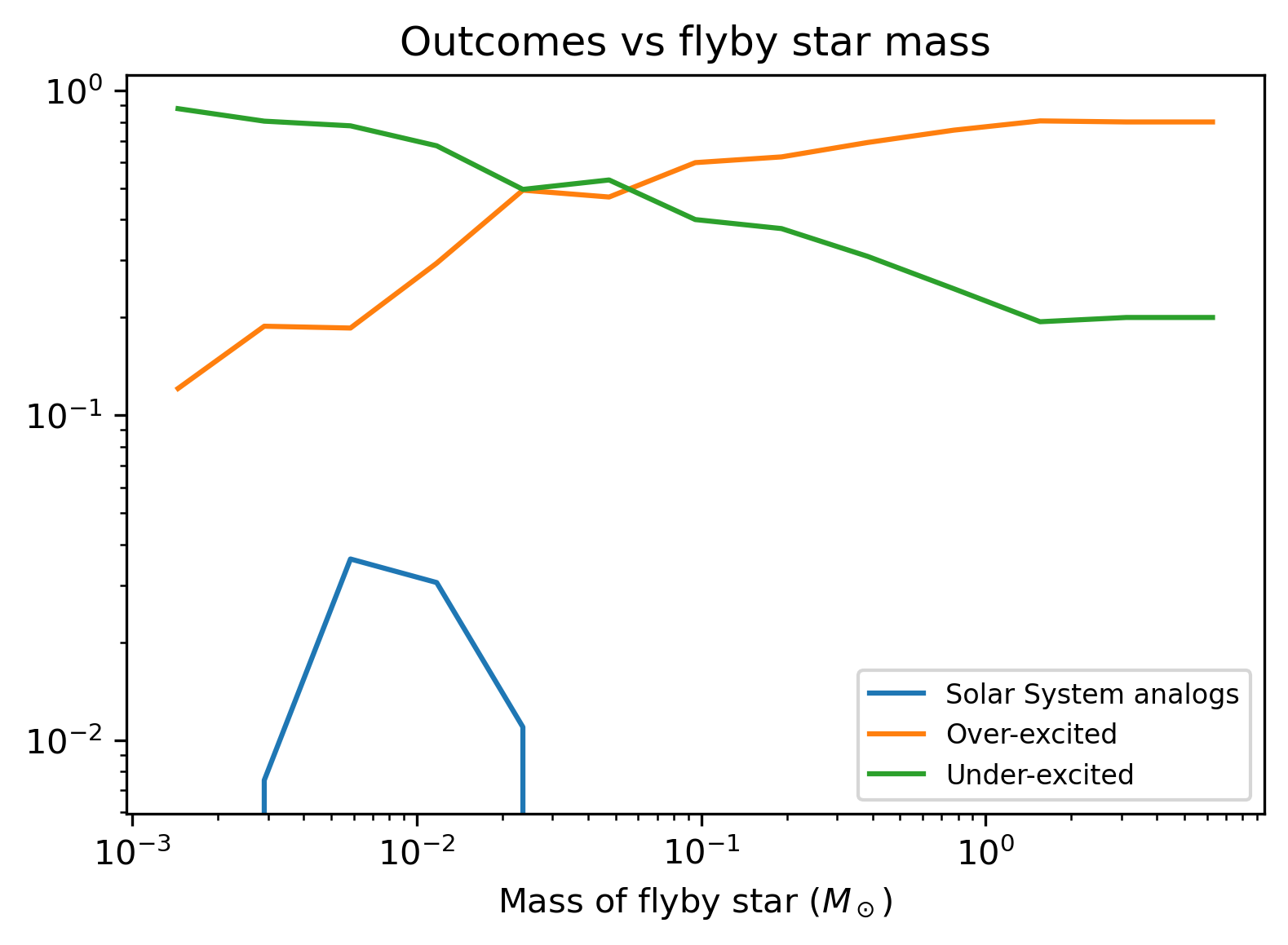}
\end{centering}
 \caption{Probability of different simulation outcomes as a function of the two most consequential flyby parameters: the closest approach distance (left) and the flyby mass (right). Solar System analogs (shown in blue) are the 20 simulations discussed in Section 3.3 that match both the giant planets and cold classical Kuiper belt. Under-excited systems are those in which no instability was triggered and the cold classicals were preserved. Over-excited systems are those in which the final system is too excited to match the Solar System, including cases in which some (or all) of the planets were stripped, or simply that no cold classical Kuiper belt objects survived.  The flyby velocity had a minimal effect. }
     \label{fig:outcomes_prob}
\end{figure*}

Our overarching goal is to map any flyby onto a distribution of outcomes with their associated probabilities. The three main outcomes are:
\begin{itemize}
\item Under-excitation. In these simulations, no instability was triggered and the cold classical Kuiper belt was left minimally perturbed.  These outcomes are consistent with the Solar System, but require a different trigger for the giant planet instability (see discussion in Section 1).
\item Solar System analogs (20/3000, or 0.7\% of our sample).  These are the cases that match the giant planets' orbits, retain the cold classical Kuiper belt, and do not strongly contaminate the hot dynamical region with objects originally within the cold classical population. These are the outcomes for which the flyby itself played the central role in triggering the instability. 
\item Over-excitation. In these cases, the final system was over-excited or under-populated.  In very violent cases, the flyby stripped all of the planets (4.2\% of simulations with impact parameter $b < 150$~au) or left a single survivor (4.9\%).  In other cases, the flyby triggered a dynamical instability that left the planets on orbits that were too excited (as measured by their $AMD$), or simply led to the loss or over-excitation of the cold classicals. These outcomes are inconsistent with the Solar System because the dynamical devastation cannot be undone.
\end{itemize}

The most important factors in determining the outcome of a flyby are the mass and closest approach. Figure~\ref{fig:outcomes_prob} shows the fraction of simulations that produced the different Solar System outcomes as a function of flyby parameters.  Only flybys that pass within $\sim 100$~au of the Sun triggered dynamical instability. Not a single simulation with a closest approach past 120 au lost a planet.  Fig.~\ref{fig:outcomes_prob} shows that the probability of over-exciting the Solar System drops drastically past $\sim 50$~au.  There is no clear dependence of the success rate on the flyby speed within the range that we tested, apart from the fact that slower flybys correlate with closer approaches (due to gravitational focusing).  Solar System analogs are only produced for a narrow range of flyby masses (Fig.~\ref{fig:outcomes_prob}).  All of the 20 successful simulations underwent a flyby with an object between $3 \mjup$ and $28 \mjup$ (median of $10.5 \mjup$), in the free-floating planet to brown dwarf mass range.  It is clearly encounters with free-floating planetary-mass objects and low-mass brown dwarfs that were the most successful in generating a Solar System-matching flyby-driven instability.  And all of the successful cases had a closest approach of less than 18 au, such that the flyby directly perturbed the system.

The main goal of this analysis is to understand flyby outcomes well enough to devise a recipe that we can implement into Monte Carlo simulations of stellar birth clusters in the next section.  For simplicity, and because of its strong scaling (see Fig.~\ref{fig:outcomes_prob}), we quantified the balance between under- and over-excitation focusing mainly on the closest approach distance.  All successful outcomes had flyby closest approaches of less than $\sim 20$~au and masses below $\sim 30 \mjup$.  Within our sample of 3000 simulations, 357 had closest approaches $r_{close} < 20~\rm{au}$ and masses $3 \mjup < M_\star < 30 \mjup$.\footnote{\cite{li15,li16} found $\sim 50$~au as the approximate cutoff for flybys that were likely to catastrophically disrupt the Solar System. Our results are in broad agreement, although it is clear that the flyby mass plays a key role, and extending to the substellar regime opens the possibility of safe, closer flybys.} Of these, 111 (31.1\%) did not undergo an instability.  These simulations are clearly not consistent with a flyby-triggered instability. Another 89 (24.9\%) simulations finished with zero, one or two planets.  These are likewise inconsistent with a flyby-driven scenario. The remaining 157 simulations (44\%) are broadly consistent with the Solar System in that they underwent a flyby-driven instability, and finished with three or four giant planets.  It's worth keeping in mind that only 20 of these simulations (5.6\% of the sample, but 12.7\% among the `broadly consistent' simulations) were classified as successes according to our analysis in Section 3.  Those 20 simulations essentially represent the fraction of instabilities that quantitatively match the Solar System's orbital architecture, as discussed in Section 3.  A plethora of previous studies has shown that only a fraction (typically $\sim 10\%$) of instability simulations actually match the Solar System in terms of the actual orbits of the giant planets~\citep[e.g.][]{tsiganis05,morby07,nesvorny12,batygin12b,kaib16,clement21a}. In that sense, we can count the full grouping of 157 (44\%) of simulations with $r_{close} < 20 \rm{au}$ and $3 \mjup < M_\star < 30 \mjup$ as being potentially successful, rather than just the 20 (5.6\%) that provided a concrete match. Yet despite the existence of other flyby-driven instabilities outside this box in parameter space (in particular, for higher flyby masses), we decided to remain within the boundaries set by those simulations that provided a quantitative match to the outer Solar System.  Within this box, 105 of the 357 simulations (29.4\%) did not undergo an instability and also retained the vast majority of the cold classical Kuiper belt on cold orbits -- those are considered `under-excited'.  The remaining 26.6\% of simulations are considered to over-excite the system.

\subsection{Monte Carlo stellar cluster flyby modeling}

We built a simple framework to sample stellar flybys in a Monte Carlo fashion based on stellar cluster parameters.  Our approach is based on that of \cite{zink20}, which itself was inspired by \cite{heisler86}.  First, we define the main cluster parameters: the stellar density $\eta$ (assumed to be constant in time) and the cluster lifetime $T$. We fix the stellar velocity distribution $\langle v \rangle$ to be 1 km/s~\citep{lada03}. We define a timestep $\Delta t$ (we used $\Delta t = 10^{3-4}$ years and did not see any difference). 

\begin{figure}
	\includegraphics[width=0.9\columnwidth]{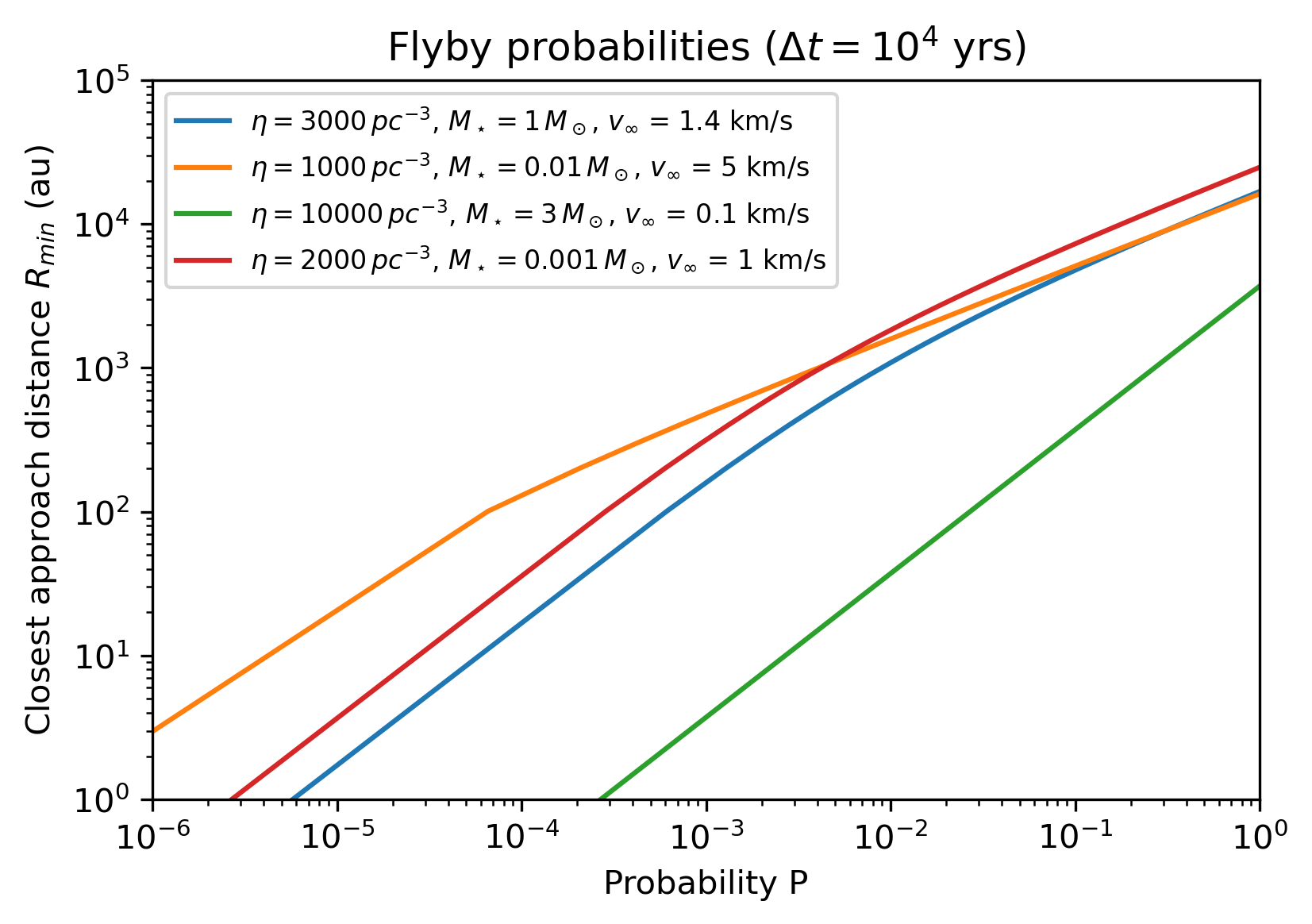}
 \caption{Probability distributions of flybys for stellar clusters with specified parameters, for a simulation timestep of $\Delta t = 10^4$ years.  See Section 4.2 for details.}
    \label{fig:flyby_prob}
\end{figure}

For each timestep, we draw a mass from the \cite{maschberger13} initial mass function between $1 \mjup$ and $10 \msun$ (as in the simulations presented above). We draw a velocity $\vinf$ from a Maxwell-Boltzmann distribution with a scale parameter of $\sqrt2 \langle v \rangle$~\citep{binney87}.  Next, we draw a random number $P$ between zero and 1.  $P$ represents the probability of an encounter within a given radius $R_{min}$ within a given timestep, and is defined as $P = \eta \, \vinf \, \pi \, R_{min}^2 \, (1 + \Theta) \, \Delta t$. $\Theta$ is the (Safronov) gravitational focusing factor, the square of the ratio of the escape speed from the Sun at $R_{min}$ to $\vinf$. We solve for $R_{min}$, which is the closest approach distance of the flyby for that specific timestep.  Figure~\ref{fig:flyby_prob} shows the probability  of an encounter within a given radius during a given timestep scales with different cluster parameters, for a fixed value of $\Delta t$.  Note that, with this approach, the probability of an encounter within a given distance depends on $\Delta t$: the smaller the timestep, the lower the probability.  Yet, over the course of a simulation (which typically lasts $10^7$ years), we carefully verified that the median closest encounter is independent of the chosen timestep. In practice, we only account for encounters with $b < 1000$~au, since that is the limit of our simulations and we have seen clearly that flybys past 100~au have no discernible effect on the planets' orbits or the cold classical Kuiper belt.

For each flyby, we evaluate the outcome based on the analysis presented in Section 4.1.  If the closest approach $R_{min}$ is greater than 100 au, then the system is assumed to be under-excited (which was the outcome in all but $\sim 1 \%$ of such simulations).  If $R_{min} = 20-100$~au, then there is an 59/41 split between under- and over-excitation, which was the outcome integrated over all flyby masses (see Fig.~\ref{fig:outcomes_prob}).  For $R_{min}<20$~au and flyby masses between 3 and $30 \mjup$, we assign a 44\% probability of a Solar System-like outcome, a 29.4\% probability of under-excitation and a 26.6\% probability of over-excitation (see Section 4.1).  Following our simulation results, we assign a 100\% probability of over-excitation for more massive flybys within 20 au, and a 76/24 split between under- and over-excitation for less massive flybys within 20 au.

We evaluate the overall outcome of each Monte Carlo realization using the list of flybys and the associated excitation probabilities. We assume that the system starts in an under-excited state similar to our simulations. We loop through each Monte Carlo flyby in sequential (time) order, calculating the probability of the Solar System being over-excited, under-excited, or producing a potential Solar System analog system. We use a random number generator to determine the outcome based on the probabilities for that specific flyby.  If the outcome is `over-excited' then it's over, and that is the final state of the entire realization.  If the outcome is `under-excited' then there is no change to whatever the previous system state was.  If the outcome is `Solar System analog,' then an under-excited system state changes to that value, and stays that way unless it gets over-excited in a later flyby.  

\begin{figure}
\begin{centering}
	\includegraphics[width=0.9\columnwidth]{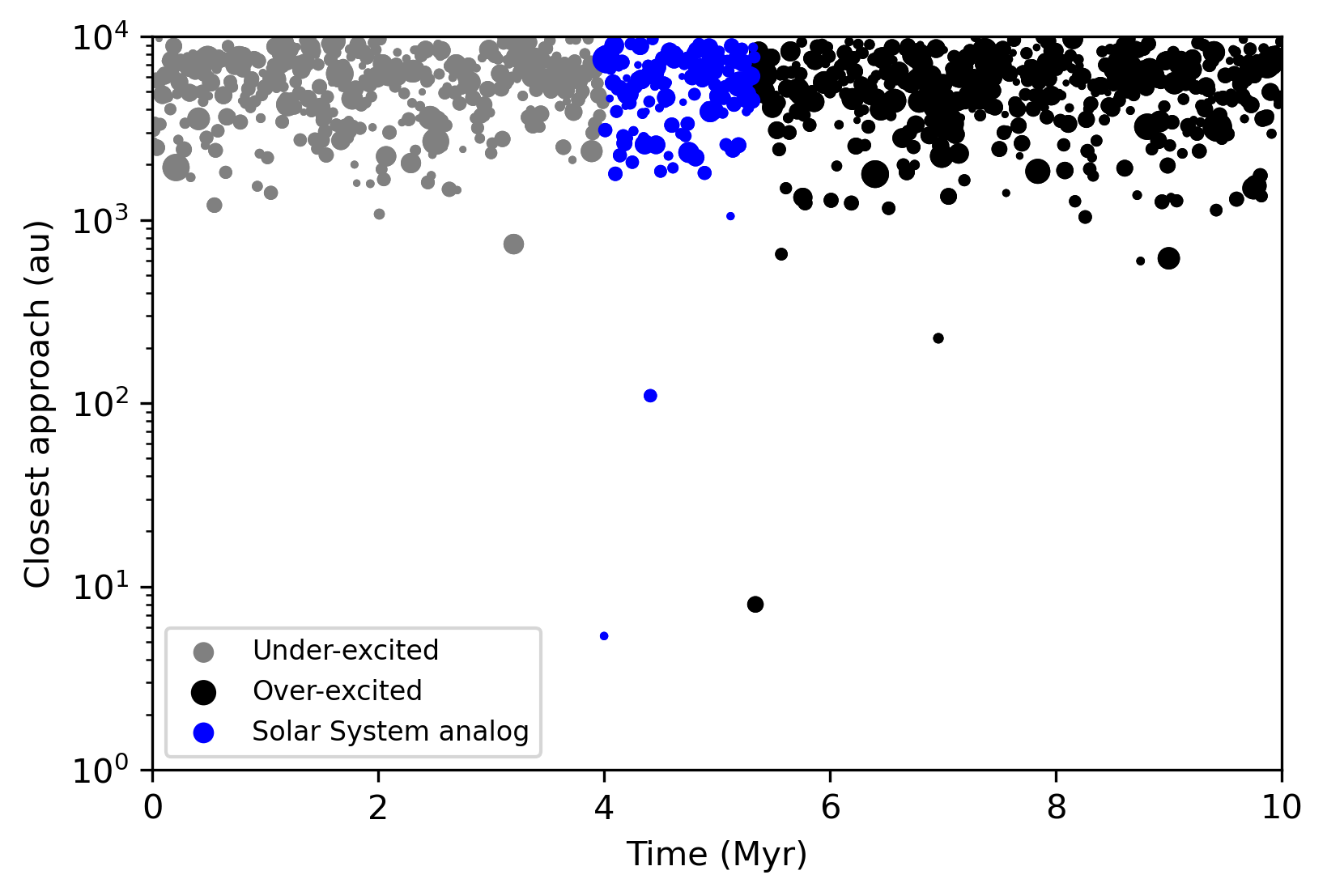}
\end{centering}
 \caption{A sample evolution of our Monte Carlo flyby code that shows the close encounters experienced by the Sun during the 10 Myr lifetime of the cluster.  Each symbol refers to a single flyby with a size that scales with the square root of the flyby mass.  Here, we fixed the cluster lifetime $T$ at 10 Myr and the stellar density $\eta$ at $8000 ~{\rm pc^{-3}}$.  Before 4.3 Myr, none of the encounters was close enough to excite the system. At that point, the flyby of a $5.5 \mjup$ free-floating planet at 5.34 au excited the system to a level consistent with a flyby-triggered instability. However, at 7.2 Myr the flyby of a $0.25 \msun$ star at 8 au over-excited the system, rendering it incompatible with the Solar System.}
    \label{fig:cluster_examples}
\end{figure}

Figure~\ref{fig:cluster_examples} shows an example Monte Carlo realization that illustrates the different outcomes. For the first several million years, there were no close encounters closer than $\sim 500$~au, and the system remained in an under-excited state.  A significant fraction of realizations behave similarly, never undergoing a close-enough flyby and remaining under-excited.  At 4 Myr, the Sun underwent the flyby of a $5.5 \mjup$ free-floating planet with a close approach of 5.4 au, which triggered a dynamical instability without overexciting the system.  For the next few million years, the system remained in a state consistent with a flyby-driven instability.  Then, at 5.34 Myr, the flyby of a $0.25 \msun$ brown dwarf at 8 au irreversibly over-excited the system.

We used our Monte Carlo code to determine the probability of different outcomes as a function of cluster parameters.  The most important parameter that we tested is the product of the stellar density $\eta$ and the cluster lifetime $T$~\citep[referred to as \Chi \, by][]{batygin20}. This parameter governs the number and importance of stellar flybys, and is a simplification since the true quantity of interest is the integral of the $\eta(t) dt$ over the cluster lifetime, keeping in mind that the local stellar density evolves substantially both as a star orbits within the cluster and as the cluster itself evolves on a timesacle of 10-100 Myr~\citep[e.g.][]{adams01,portegieszwart09}. In our Monte Carlo simulations, we vary $\eta T$ between $10^3$ and $10^6$ Myr pc$^{-3}$. Given the probability scaling for close encounters discussed above and shown in Fig.~\ref{fig:flyby_prob}, our Monte Carlo simulations are agnostic as to whether it is the duration of the simulations or the stellar density that varies.  For instance $\eta T = 10^4$ Myr pc$^{-3}$ could correspond to the Sun spending 100 Myr in a cluster with a number density of 100 stars per cubic parsec~\citep[a commonly-found value in the literature; e.g.][]{lada03}, or 3 Myr in a cluster with a number density of $\sim 3000$ stars per cubic parsec. Observations of embedded star clusters find a range of many orders of magnitude in the mass density of clusters~\citep[e.g.][]{lada03,portegieszwart09,adams10}, as well as two different evolutionary tracks for such clusters~\citep[e.g.][]{pfalzner09,portegieszwart10}. 

Since our stellar initial mass function extends down to the substellar range, any specified value of $\eta T$ would include fewer actual (main sequence) stars than in a traditional study that only considers stars.  Within our chosen initial mass function, $\sim 40\%$ of objects have masses lower than $0.1 \msun$. Accounting for this, the $\eta T$ values in our study are almost a factor of two higher than for corresponding studies that neglect brown dwarfs and free-floating planets. To account for this, we use $\eta T_\star$ to refer to $\eta T$ values in the literature that have a cutoff at the substellar mass limit of $\sim 0.08 \msun= 80 \mjup$, keeping in mind that $\eta T \approx 2 \eta T_\star$.  This may seem surprising, since it is well known that brown dwarfs are relatively few in number compared with M dwarfs~\citep[e.g.][]{luhman03,hennebelle08,bate18,haugbolle18}.  However, this is compensated by the fact that our low-mass limit is so much smaller than the substellar mass limit (a factor of 80 in mass).

\begin{figure}
	\includegraphics[width=0.9\columnwidth]{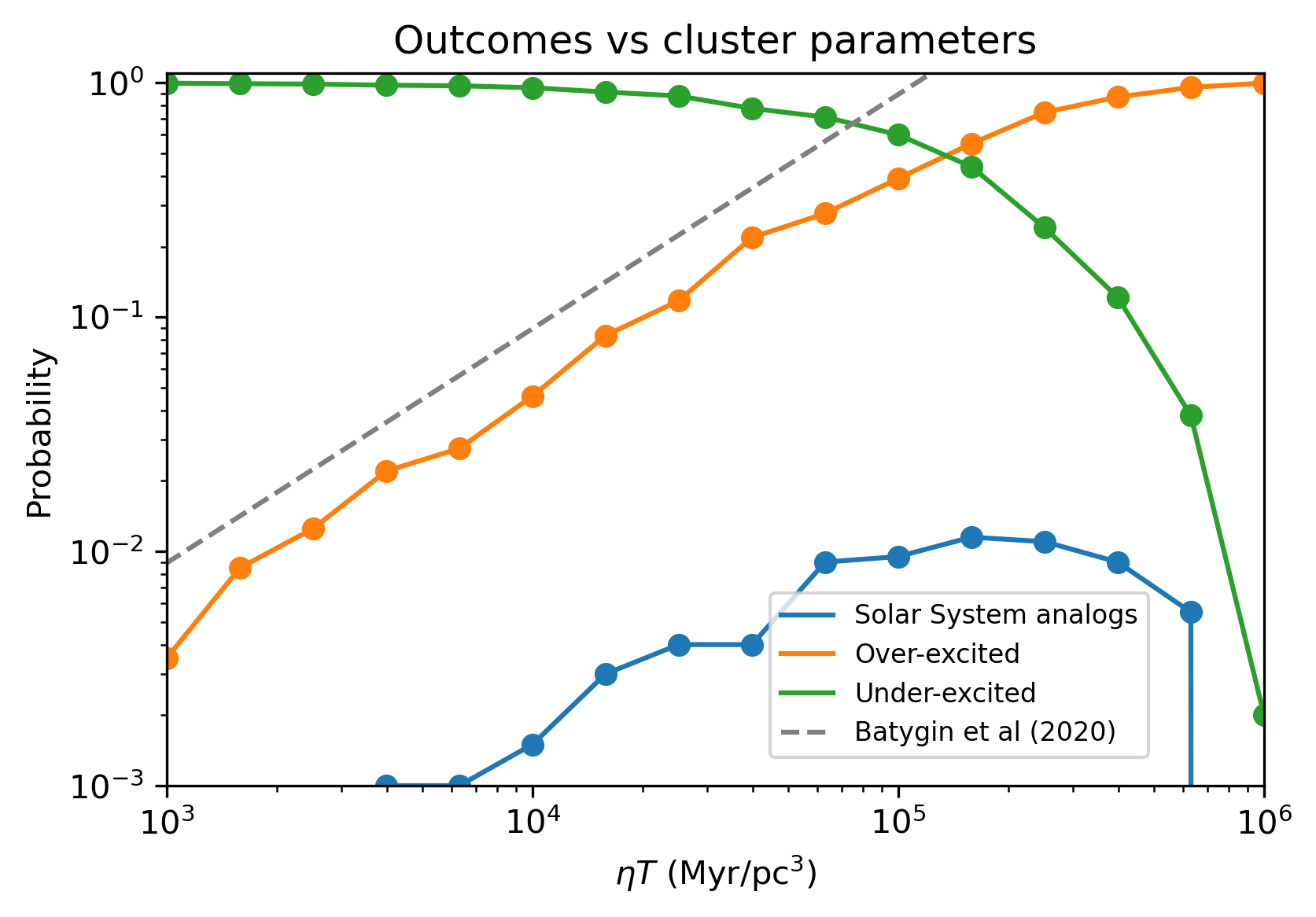}
 \caption{Fractional distribution of the outcomes of our Monte Carlo simulations.  For each value of $\eta T$, we ran 1000 realizations.  We shifted the results of \cite{batygin20} to higher $\eta T$ to account for the fact that $\sim 40\%$ of the stars in our sample have masses below the substellar cutoff. }
    \label{fig:cluster_MC}
\end{figure}

Figure~\ref{fig:cluster_MC} shows that the balance between under- and over-excitation of the Solar System is a strong function of $\eta T$. This is no surprise, as $\eta T$ essentially measures the cumulative strength of stellar flybys. The crossover point between the dominance of under- vs. over-excitation is at $\eta T \approx 10^5$~Myr pc$^{-3}$, which corresponds to $\eta T_\star \approx 5 \times 10^4$~Myr pc$^{-3}$ in previous studies. \cite{nesvorny23} showed that, in order to reproduce the radial distribution of distant TNOs, the Sun must have formed in a cluster with $\eta T_\star \gtrsim 10^4$~Myr pc$^{-3}$.  \cite{batygin20} showed that  the probability of disrupting the cold classical Kuiper belt increases with $\eta T$, and their curve tracks quite close to our own derived probability for over-excitation (dashed line in Fig.~\ref{fig:cluster_MC}). It is notable that our Monte Carlo-generated clusters appear to perturb the cold classicals somewhat less than in those of \cite{batygin20}. We suspect that this difference is again linked with the mass range of stars, which extend all the way up to $20 \msun$ for \cite{batygin20}. 

The probability that the Solar System was shaped by a flyby-driven dynamical instability increases with $\eta T$, but more slowly than the probability of over-excitation (see Fig.~\ref{fig:cluster_MC}).  The Solar System analog rate increased from $\lesssim 10^{-3}$ for $\eta T < 10^4$~Myr pc$^{-3}$ to a plateau at $\sim 1\%$ for $\eta T$ between roughly $6 \times 10^4$ and a few $\times 10^5$~Myr pc$^{-3}$.  At higher values of $\eta T$, the Solar System analog rate drops steeply due to the predominance of very strong encounters that over-excite the system~\citep[for simulations showing how encounters within a cluster can disrupt planetary systems, see, for example][]{malmberg11,li15,stock20,rickman23}.  

The low abundance of free-floating planets and brown dwarfs is a main limiting factor for a flyby-driven instability.  Observations of young stellar clusters find an over-abundance of objects in that mass range relative to predictions~\citep{miretroig22,defurio25}, although the magnitude of the overabundance remains poorly characterized~\citep[see][]{luhman25}.  Although they probe a different galactic population than direct imaging, microlensing surveys also find an overabundance of free-floating planets relative to stars~\citep{sumi23}.  

We tested the effect of increasing the number of low-mass brown dwarfs and free-floating planets.  For our standard \cite{maschberger13} stellar initial mass function between $1 \mjup$ and $10 \msun$, the median mass is $\sim 0.07 \msun$, and 31.8\% of all objects have masses below $30 \mjup$.  To crudely mimic an increase in the abundance of free-floating planets and low-mass brown dwarfs, we hold the total number of objects with $M > 30 \mjup$ fixed, but increase the number of objects with $1 \mjup < M < 30 \mjup$ by a factor $F$. For $F=1$, our IMF is simply the \cite{maschberger13} one.  For the maximum value of $F = 10$, i.e., a ten-fold increase in the number of low-mass brown dwarfs and free-floating planets, there are almost five times more objects with $M < 30 \mjup$ relative to those with $M > 30 \mjup$. In practice, $F$ is implemented in our Monte Carlo code in two places: it skews the choice of the stellar mass during a given timestep, and increases the total stellar density by $0.318 F$ (because 0.318 is the fraction of objects below $30 \mjup$).  A high value of $F$ can increase the effective stellar density considerably, which yields statistically closer flybys when the abundance of free-floating planets is increased (see Fig.~\ref{fig:flyby_prob}).  For this test, we fixed the parameter $\eta T = 7 \times 10^4$~Myr pc$^{-3}$ -- this corresponds to $\eta T_\star \approx 3.5 \times 10^4$~Myr pc$^{-3}$ and is consistent with constraints from the cold classical Kuiper belt and the radial distribution of TNOs~\citep[e.g.,][]{batygin20,nesvorny23}.

\begin{figure}
	\includegraphics[width=0.9\columnwidth]{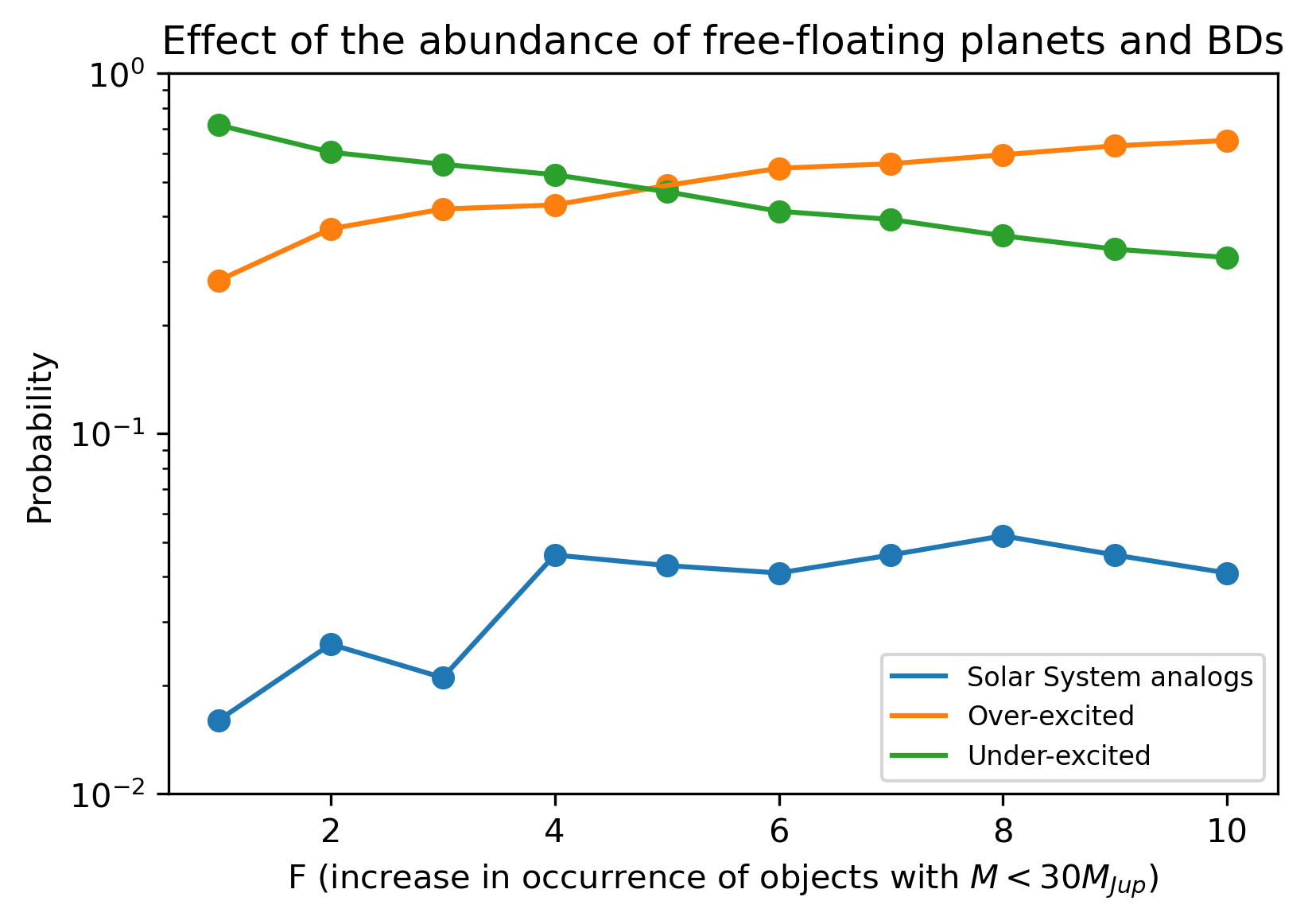}
 \caption{Outcomes of our Monte Carlo simulations as a function of $F$, the fractional increase in the number of free-floating planets and low-mass brown dwarfs (with $1 \mjup < M < 30 \mjup$) relative to stars.  Here, we fixed $\eta T = 7 \times 10^4$~Myr pc$^{-3}$. }
    \label{fig:cluster_FFP}
\end{figure}

Figure~\ref{fig:cluster_FFP} shows that the probability of a flyby-driven dynamical instability in the Solar System increases with $F$.  At $F=1$, the rate is $\sim$1\%, matching the peak from Fig.~\ref{fig:cluster_MC}. The Solar System analog rate increases quickly to $\sim$5\% for $F \geq 4$, and plateaus out to higher values of $F$, emphasizing the importance of even a modest underestimation of the population of free-floating planets and low-mass brown dwarfs. The probability of system over-excitation also increases with $F$, simply because the number of very close flybys increases.  While not seen at the exact transition, it is interesting to note that the peak in the success rate for Solar System analogs again falls close to the transition between under- and over-excitation (as seen in Fig.~\ref{fig:cluster_MC}).

All together, our analysis suggests that the probability that the Solar System's dynamical instability was triggered by a stellar flyby was at least one percent, and $\sim$5\% if the occurrence of free-floating planets is even modestly underestimated by standard stellar initial mass functions.  

\section{Is a dynamical instability during the birth cluster phase consistent with the Oort cloud?}

One might wonder whether our setup, with a giant planet dynamical instability taking place during the Sun's birth cluster phase, is consistent with the existence and mass of the Oort cloud.  The Oort cloud, whose properties are inferred from the orbits of long-period comets on roughly parabolic orbits~\citep{oort50}, extends out to the Sun's ionization radius at about 1 pc~\citep{smoluchowski84}.  It is estimated to contain a total of roughly an Earth-mass in cometary nuclei, although there is a factor of a few uncertainty on this value~\citep[see discussion in][]{kaibvolk24}.  

The Oort cloud is thought to have been populated by planetesimals scattered outward by the giant planets, whose orbits were torqued by passing stars or gradients in the local gravitational field~\citep[e.g.][]{heisler86,duncan87,fernandez97,wiegert99,brasser06,kaib08,dones15,portegies21}. While the Sun was in its birth cluster, its ionization radius was about an order of magnitude smaller than it is today~\citep[][this is the reason for the $10^4$ au ejection radius used in our simulations]{tremaine93}.  Planetesimals scattered outward by the giant planets during this phase could only be trapped within the inner Oort cloud, with semimajor axes of $\sim 1,000-5,000$ au~\citep[e.g.][]{saillenfest20}.  The outer Oort cloud can only have been populated after the Sun left its cluster.  The question becomes, are there enough planetesimals to scatter into the Oort cloud after the Sun left its birth cluster?

\begin{figure}
	\includegraphics[width=0.9\columnwidth]{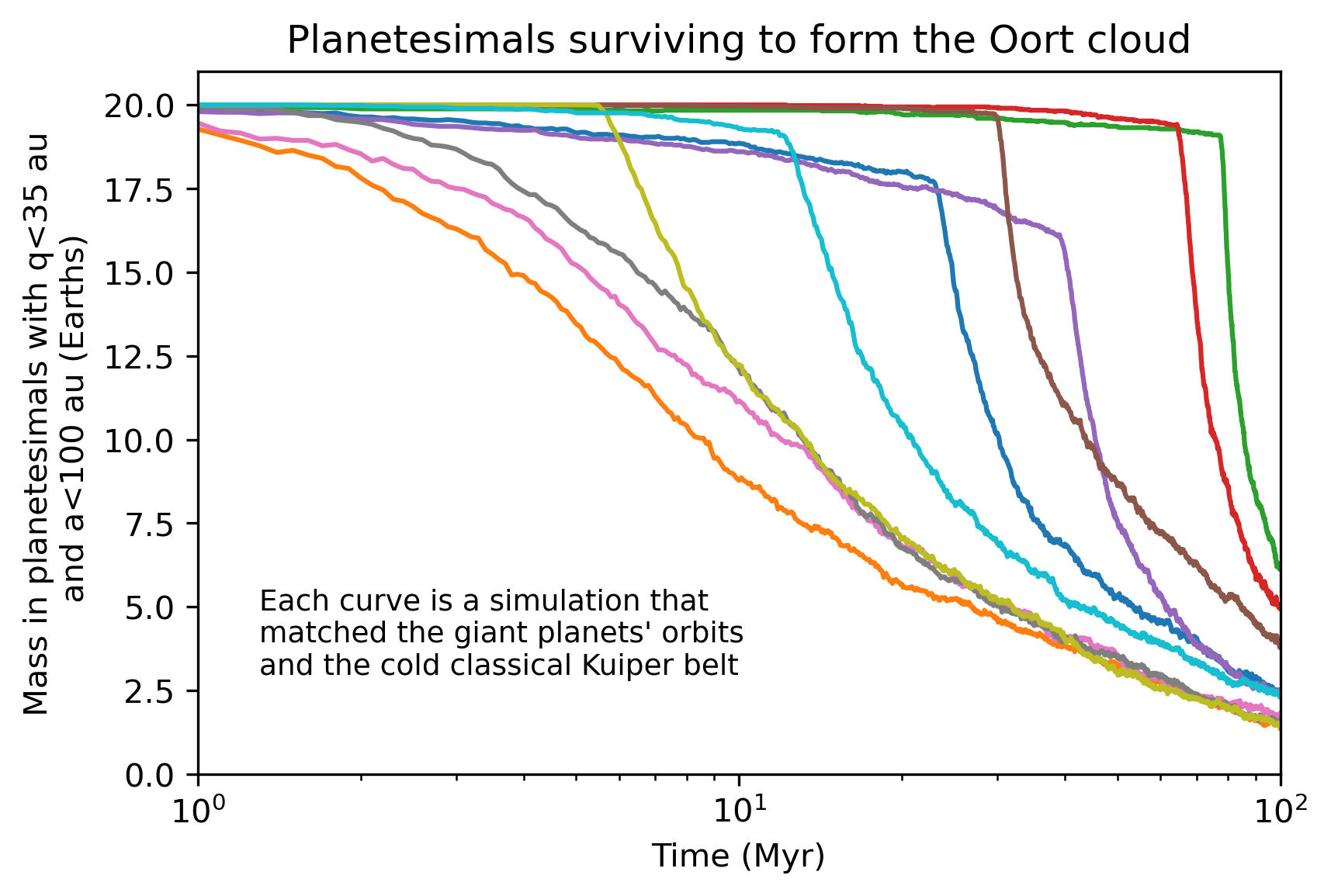}
 \caption{The mass remaining in planetesimals in ten example simulations that matched the giant planets' orbits and cold classical Kuiper belt.  We only consider planetesimals that are in dynamical contact with the planets (with perihelia $q < 35$ au and with semimajor axes small enough not to be affected by the cluster tide ($a < 100$ au).  The simulations had a wide range in instability timescales, as can easily be deduced by the time when the number of planetesimals starts to drop sharply. }
    \label{fig:OC}
\end{figure}

Figure~\ref{fig:OC} shows how the total mass in planetesimals interacting with the giant planets drops as a function of time in ten different simulations that matched the giant planets and cold classical Kuiper belt. Here, we have only included planetesimals with semimajor axes smaller than 100 au, to ensure that they are immune to torques from the tidal effects of the cluster itself~\citep{tremaine93}.  The examples cover a wide range in instability timescales, which are clearly identified in the Figure as the time when the planetesimal mass starts to drops sharply.  To interpret this plot in the context of the formation of the Oort cloud, we need to know when the Sun left its birth cluster.  Constraints come from the existence of radioactive nuclides in meteorites that were produced in massive stars whose lifetimes are known~\citep[see discussion in][]{arakawa23}. Yet not all stars within a given star-forming association are co-eval, and the relevant time zero in our analysis is after the dispersal of the Sun's protoplanetary disk, which lasted about 5 Myr~\citep{wang17b,hunt22}. For simplicity, let's adopt a commonly-assumed value of 10 Myr for the Sun's residence time within the cluster.  We can see from Fig.~\ref{fig:OC} that more than half of the total planetesimal mass remains after 10 Myr in all of our examples, including simulations in which the instability was triggered instantaneously by the flyby. Simulations of Oort cloud formation typically find that $\sim 5-10\%$ of scattered planetesimals are trapped in the cloud when neglecting the effect of a cluster environment~\citep[e.g.][]{kaib08,dones15}.  This implies that the simulations from Fig.~\ref{fig:OC} would indeed form (outer) Oort clouds with at least $\sim 0.5-1 \mearth$ (assuming a cluster dispersal time of $\sim 10$~Myr), which is consistent with observations. \cite{izidoro25} found a similar result, that an early dynamical instability during the cluster phase is consistent with the present-day Oort cloud.  This might be problematic if the Sun remained in the birth cluster for $\sim 100$ Myr, but for shorter timescales there is no problem explaining the outer Oort cloud~\citep[see also][]{portegies25}.

There exists an alternate scenario in which the Sun's Oort cloud was captured from planetesimals scattered outward by other stars within the Sun's birth cluster~\citep{levison10}.  During the cluster's dispersal, the Sun would have captured a fraction of these scattered planetesimals.  This scenario is entirely plausible, and total mass in the Sun's Oort cloud would depend on the abundance of planetesimals scattered outward by the ensemble of stars within the cluster (and therefore on the demographics of planetesimal-scattering planets), as well as on the timescale of cluster dispersal.  

It is also possible that Planet 9 -- the putative $\sim 5 \mearth$ planet on an orbit a few hundred au in size~\citep{batygin16,batygin19} -- could have played a role in scattering planetesimals to the outer Oort cloud.

\section{Summary and discussion}

In this paper, we addressed the question of whether a stellar or substellar flyby during the Sun's birth cluster phase could have triggered the giant planet dynamical instability.  We ran 3000 dynamical simulations of flybys of the young Solar System, assuming that the giant planets were in a multi-resonant configuration that would remain stable if left unperturbed.  Our simulated flybys had impact parameters from 1 to 1000~au, velocities of up to 5 km/s, and masses from $1 \mjup$ to $10 \msun$.  We found clear trends in the outcomes of flybys, with very strong flybys exciting the system beyond the current level, very weak flybys having no discernible effect, and an intermediate range in which a flyby could trigger a dynamical instability without leaving the system over-excited (see Section 3).  

Simulations that reproduce the outer Solar System -- both matching the giant planets' orbits (Fig.~\ref{fig:finalorbits}) and preserving the cold classical Kuiper belt (Fig.~\ref{fig:outerSS}) -- involve the flyby of a substellar object (mass range $3-30 \mjup$) crossing directly within the Solar System, with a closest approach smaller than 20 au (Fig~\ref{fig:outcomes_prob}).  Within that parameter range, 5.6\% of simulations quantitatively matched the giant planets and Kuiper belt, although a larger fraction (44\%) with less successful outcomes could nonetheless be interpreted as having undergone a dynamical instability broadly consistent with the one suffered by the young Solar System (see discussion in Section 4.1).

We performed Monte Carlo simulations of the flybys that may have been suffered by the young Sun in its birth cluster, calibrated to our N-body simulations.  As in previous work, the cluster was parameterized by $\eta T$, the product of the stellar density $\eta$ and the cluster lifetime $T$; simply put, higher values of $\eta T$ indicate closer and stronger stellar encounters. The crossover between under- and over-excitation of the young Solar System happened for $\eta T \approx 10^5$~Myr pc$^{-3}$, which corresponds to $\eta T_\star \approx 5 \times 10^4$~Myr pc$^{-3}$ in previous studies to account for flybys in our work that fall below the substellar boundary. The highest probability that the Solar System's instability was triggered by a flyby was roughly 1\%, for $\eta T$ values near the under-/over-excitation boundary (see Fig.~\ref{fig:cluster_MC}).  When we increased the relative abundance of free-floating planets and low-mass brown dwarfs (with $M = 1-30 \mjup$) by a factor of a few~\citep[as suggested by observational studies;][]{miretroig22,defurio25}, the probability of a flyby-triggered instability increased to $\sim$5\% (see Fig.~\ref{fig:cluster_FFP}). 

\begin{figure}
	\includegraphics[width=0.9\columnwidth]{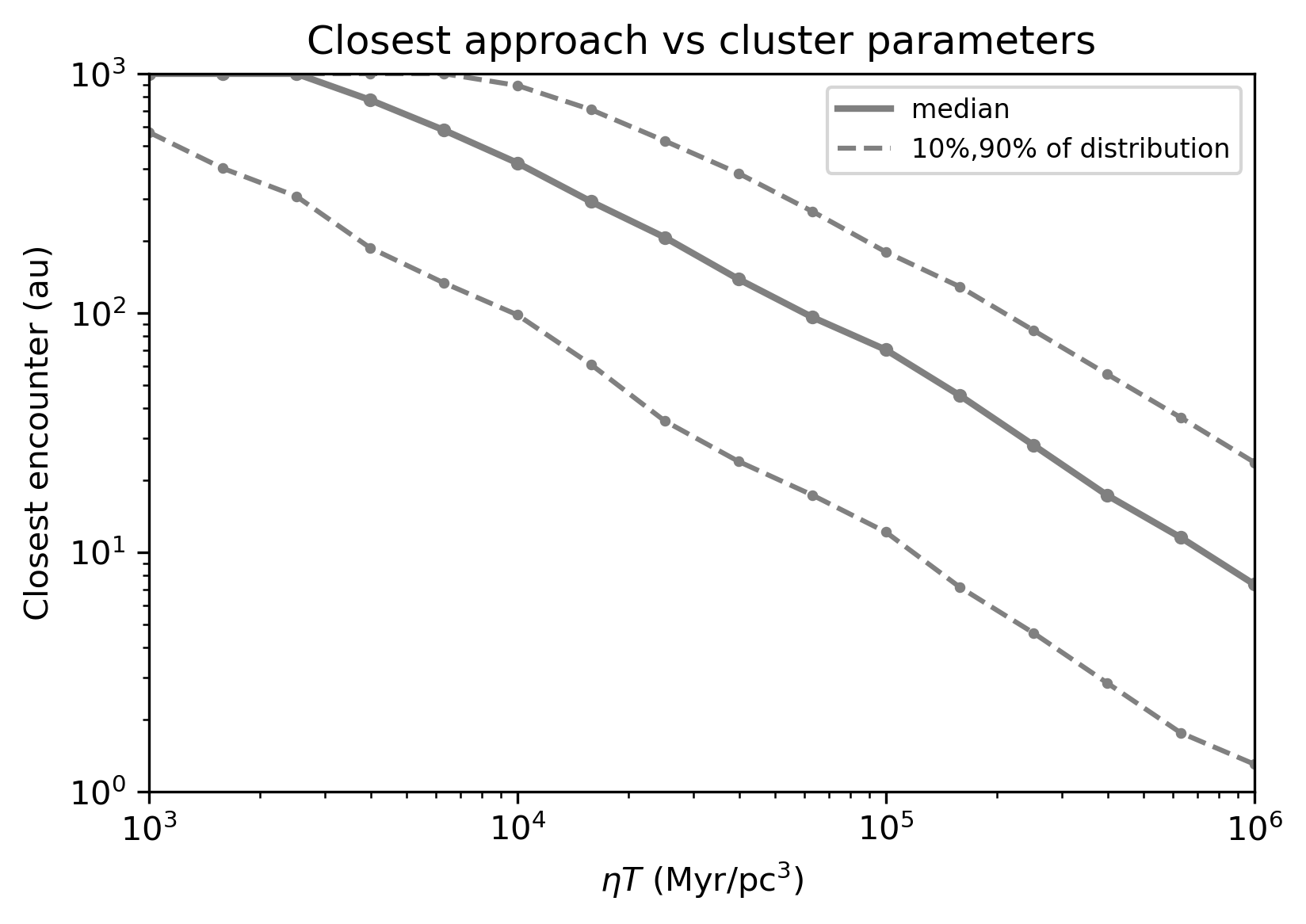}
 \caption{Distribution of the closest encounter felt by the Sun as a function of $\eta T$, using the same calculations as from Fig.~\ref{fig:cluster_MC}.  The figure shows the median closest encounter for 1000 Monte Carlo realizations at each value of $\eta T$, as well as the 10\% and 90\% range of the distribution. }
    \label{fig:cluster_rclose}
\end{figure}

The key factor determining the outcome of the flybys suffered by the young Sun is the single strongest (sub)stellar encounter. This is usually (but not always) the closest encounter -- see for example the case illustrated in Fig.~\ref{fig:cluster_examples}.  Figure~\ref{fig:cluster_rclose} shows the median closest flyby as a function of $\eta T$.  Here, $\eta T$ is the same as $\eta T_\star$ in previous studies~\citep[e.g.][]{batygin20,arakawa23,nesvorny23,siraj25} because we are simply measuring the closest approach, not its impact on the planetary system.  For $\eta T = 10^4$~Myr pc$^{-3}$, at the low end of generally accepted values, the median closest flyby is 424 au, but the 10\%-90\% range extends from 103~au to 870 au.  At $\eta T = 10^5$~Myr pc$^{-3}$, representing the high end of plausible values, the median closest flyby was 95 au, with a 10\%-90\% range extending from 14-270 au.  

The distribution of the closest flybys from Fig.~\ref{fig:cluster_rclose} offers two important messages.  First, given the wide range of closest encounter distances for any given value of $\eta T$, it is hard to imagine constraining $\eta T$ overly precisely, at least using dynamical constraints.  Previous work by \cite{nesvorny23} showed that to match the radial distribution of TNOs, the Sun's birth cluster must have had $\eta T_\star \gtrsim 10^4$~Myr pc$^{-3}$ (recall that $\eta T \approx 2 \eta T_\star$).  Meanwhile, \cite{batygin20} showed that the cold classical Kuiper belt is unlikely to have survived for $\eta T_\star \gtrsim 5 \times 10^4$~Myr pc$^{-3}$ (corresponding to $\eta T \approx 10^5$~Myr pc$^{-3}$ in Fig.~\ref{fig:cluster_MC}). In contrast, \cite{siraj25} used the inclination distribution of sednoids to propose that $\eta T_\star \lesssim 5 \times 10^3$~Myr pc$^{-3}$. This apparent contradiction may be resolved if each study constrains $\eta T_\star$ at a different time -- or, rather, the integrated $\eta(t) dt$ from a different time until the present day. While the cold classical Kuiper belt population must have formed during the Sun's protoplanetary disk phase~\citep[e.g.][]{li25}, which only lasted perhaps 5 Myr~\citep[e.g.][]{haisch01,wang17b,hunt22}, the population of more distant Trans-Neptunian objects was scattered outward, mainly during the giant planet instability~\citep[e.g.][]{brasser13b}.  Constraints imposed by the cold classical Kuiper belt therefore extend back to an earlier era in Solar System history than those from the distribution of scattered TNOs, and the time interval between the two depends strongly on the trigger of the giant planet dynamical instability (see discussion in the Introduction).  It would make sense for constraints imposed by the cold classical Kuiper belt to derive higher values of $\eta T_\star$ that constraints related to the sednoids or the distribution of distant TNOs. The acceptable range in $\eta T_\star$ values from \cite{batygin20} for the cold classicals and \cite{nesvorny23} for the radial distribution of TNOs overlap, as do our own results regarding the giant planet instability and cold classicals.  However, those of \cite{siraj25} for the sednoids do not.  The source of this discrepancy is unclear, and could potentially indicate that there was considerable time required for the sednoids to reach their current orbits, or perhaps it is affected by the small number of sednoids available for such analysis, a problem that will likely be alleviated by upcoming discoveries made with the Vera C. Rubin Observatory's Legacy Survey of Space and Time (LSST).   Nonetheless, given the breadth in the distribution of closest encounters shown in Fig.~\ref{fig:cluster_rclose}, we believe that there remains considerable uncertainty in the actual value of $\eta T$ (or $\eta T_\star$) for the Sun's birth cluster.

Second, it is entirely within the realm of possibility that the Sun underwent a planet- or outer planetesimal disk-crossing encounter during the birth cluster phase.  The most probable object to make such a close passage would have been a free-floating planet or low-mass brown dwarf, because a) the abundance of these substellar objects is likely underestimated by standard stellar initial mass functions~\citep{miretroig22,defurio25}, and b) the flyby of such a low-mass object is less likely to have completely disrupted the Solar System. Rather, it could potentially explain features of the Solar System.  To our knowledge, \cite{brown25} were the first to consider such an event as actually having shaped the early Solar System.  Our work validates and builds on their approach.

We conclude that a substellar flyby is a viable trigger for the giant planet dynamical instability.  This represents a fourth plausible instability trigger, in addition to the gas-driven~\citep{liu22}, self-driven~\citep{ribeiro20,griveaud24}, and planetesimal-driven~\citep{tsiganis05,levison10,quarles19,kaib24b} scenarios discussed in the Introduction. However, even though an instability-triggering flyby would have happened during the cluster lifetime, within $\sim$10 Myr of the start of Solar System formation, the instability itself could potentially have been delayed by up to $\sim$100 Myr, making it hard to tell apart from a planetesimal-driven scenario (see Fig.~\ref{fig:examples}).  Roughly half of our successful simulations fall into this category, with instability times later than $\sim$10 Myr after the flyby.

The nature of free-floating planets and low-mass brown dwarfs remains poorly constrained.  They may have been ejected from their home planetary systems during their own dynamical instabilities~\citep[e.g.][]{veras12}, represent the tail end of the stellar initial mass function~\citep[e.g.][]{hennebelle08,bate18,haugbolle18}, or stellar embryos whose growth was aborted~\citep[e.g.][]{reipurth01}.  If ejection represents their dominant formation channel, then one might expect their number density within the cluster to increase as a function of time, as more and more protoplanetary disks form systems of giant planets that go unstable~\citep[e.g.][]{chatterjee08,juric08,raymond10}.  Our understanding of the importance of free-floating planets and low-mass brown dwarfs for the early evolution of planetary systems will improve along with our ability to accurately characterize the populations of substellar objects in young star-forming regions~\citep[e.g.][]{pecaut16,luhman16,kirkpatrick19,miretroig22,defurio25,luhman25}.

Our simulations are admittedly imperfect.  In our N-body simulations, we only used a single multi-resonant configuration for the pre-instability giant planets' orbits~\citep[taken from][]{kaib16}. Future work could test other configurations with different radial extents and with different resonant amplitudes -- as discussed above, resonant chains created by implementing synthetic forces in N-body simulations~\citep[e.g.,][]{nesvorny12,batygin12b,kaib16,clement21a} may be somewhat more stable than those generated in more realistic, hydrodynamical simulations~\citep[e.g.,][]{pierens14,griveaud24}.  In addition, it would be interesting to perform flyby simulations at earlier phases, taking the gaseous disk into account.  Our Monte Carlo simulations were also highly simplified.  They could be improved by using actual flybys from N-body simulations of stellar clusters that capture their substructure and mass segregation, as well as by modeling the cluster dispersal phase.

\vskip .5in
{\it Acknowledgments.} 
We thank the two anonymous referees and Editor Alessandro Morbidelli for helpful comments that improved the paper.
SNR acknowledges interesting discussions with Seth Jacobson, Hanno Rein, Garret Brown, David Nesvorny, and Herve Bouy.  NAK's contributions to this work were supported from NASA Solar System Workings grant 80NSSC24K1874, NSF CAREER award 2405121 and NASA Emerging Worlds grant 80NSSC23K0771.
SNR acknowledges funding from the Programme Nationale de Planetologie (PNP) of the INSU (CNRS), and in the framework of the Investments for the Future programme IdEx, Universite de Bordeaux/RRI ORIGINS.


\end{document}